\documentclass[5p,times,twocolumn]{elsarticle}
\usepackage{lineno,hyperref}
\usepackage{subcaption}
\usepackage{tikz}
\usepackage{pgfplotstable}
\usepackage{pgfplots}
\usepackage{multirow}
\usepackage{float}
\usepackage{textcomp}
\usepackage{amsmath,amsfonts,amssymb,bbm}	
\usepackage{multicol}

\usepackage{booktabs}
\usepackage{blindtext}

\makeatletter
\def\ps@pprintTitle{%
    \let\@oddhead\@empty
    \let\@evenhead\@empty
    \def\@oddfoot{\reset@font\hfil\thepage\hfil}
    \let\@evenfoot\@oddfoot
}
\makeatother

\usepackage{xcolor}

\usepackage[section]{placeins}

\modulolinenumbers[5]

\biboptions{authoryear}

\begin{document}

\begin{frontmatter}


\title{Complex-valued neural networks to speed-up MR Thermometry during Hyperthermia using Fourier PD and PDUNet}

\author[1,2,3,]{Rupali Khatun}

\author[4,5,7]{Soumick Chatterjee}
\author[2,3]{Christoph Bert}
\author[6]{Martin Wadepohl}
\author[2,3]{Oliver J. Ott}
\author[2,3]{Sabine Semrau}
\author[2,3]{Rainer Fietkau}
\author[4]{Andreas Nürnberger}
\author[1,2,3,8]{Udo S. Gaipl}
\author[1,2,3,8]{Benjamin Frey}


\address[1]{Translational Radiobiology, Department of Radiation Oncology, Universitätsklinikum Erlangen , Friedrich-Alexnder-Universität Erlangen-Nürnberg, Erlangen, Germany}
\address[2]{Department of Radiation Oncology, Universitätsklinikum Erlangen, Friedrich-Alexnder-Universität Erlangen-Nürnberg Erlangen, Germany}
\address[3]{Comprehensive Cancer Centre Erlangen-EMN, Erlangen, Germany}
\address[4]{Data and Knowledge Engineering Group, Faculty of Computer Science, Otto von Guericke University Magdeburg, Germany}
\address[5]{Genomics Research Centre, Human Technopole, Milan, Italy}
\address[6]{Dr Sennewald Medizintechnik GmbH, Munich, Germany}

\address[7]{R. Khatun and S. Chatterjee have Equal Contribution as joint first authors}
\address[8]{U. S. Gaipl and B. Frey have Equal Contribution as joint senior authors}

\begin{abstract}
Hyperthermia (HT) in combination with radio- and/or chemotherapy has become an accepted cancer treatment for distinct solid tumour entities. In HT, tumour tissue is exogenously heated to temperatures between 39 and 43 ℃ for 60 minutes. Temperature monitoring can be performed non-invasively using dynamic magnetic resonance imaging (MRI). 
However, the slow nature of MRI leads to motion artefacts in the images due to the movements of patients during image acquisition. By discarding parts of the data, the speed of the acquisition can be increased - known as undersampling. However, due to the invalidation of the Nyquist criterion, the acquired images might be blurry and can also produce aliasing artefacts.
The aim of this work was, therefore, to reconstruct highly undersampled MR thermometry acquisitions with better resolution and with fewer artefacts compared to conventional methods. The use of deep learning in the medical field has emerged in recent times, and various studies have shown that deep learning has the potential to solve inverse problems such as MR image reconstruction. However, most of the published work only focuses on the magnitude images, while the phase images are ignored, which are fundamental requirements for MR thermometry.
This work, for the first time, presents deep learning-based solutions for reconstructing undersampled MR thermometry data. Two different deep learning models have been employed here, the Fourier Primal-Dual network and the Fourier Primal-Dual UNet, to reconstruct highly undersampled complex images of MR thermometry. MR images of 44 patients with different sarcoma types who received HT treatment in combination with radiotherapy and/or chemotherapy were used in this study. The method reduced the temperature difference between the undersampled MRIs and the fully sampled MRIs from 1.3 ℃ to 0.6 ℃ in full volume and 0.49 ℃ to 0.06 ℃ in the tumour region for an acceleration factor of 10. 

\end{abstract}

\begin{keyword}
\texttt{Hyperthermia}\sep MRI \sep  MR Image Reconstruction\sep Deep Learning \sep  Undersampled MRI, Undersampled \sep MR Reconstruction \sep Complex Image \sep MR Thermometry
\end{keyword}

\end{frontmatter}


\section{Introduction}
\noindent Hyperthermia (HT) has become one of the well-accepted cancer treatments in combination with radio- and/or chemotherapy. In HT, tumour tissue is exogenously heated to temperatures between 39 and 44 ℃ for 60 minutes to sensitise tumour cells for chemo- and/or radiotherapy \citep{cihoric2015hyperthermia,datta2019efficacy,van2002heating,kok2020locoregional}.
Temperature monitoring is an important part of quality-controlled HT and can be performed non-invasively by Magnetic Resonance Imaging (MRI). However, a major challenge is that MRI is inherently slow during several traditional sequences \citep{chatterjee2022reconresnet}. Consequently, the scan time for high-resolution imaging is long, which reduces temporal resolution. Longer scan times can also lead to an increase in motion artefacts due to patient movements during image acquisition. The speed of image acquisition can be increased by discarding parts of the data, known as undersampling \citep{chatterjee2022reconresnet}. However, this leads to blurriness and can also produce aliasing artefacts due to invalidation of the Nyquist criterion \citep{nyquist1928certain, shannon1949communication}.
Hence, MR image reconstruction and reduction of motion artefacts are in high demand. This work aims to reconstruct highly undersampled MR thermometry acquisitions of patients with sarcoma with better resolution and with fewer artefacts compared to conventional techniques such as compressed sensing. The use of deep learning in the medical field is spreading, including for undersampled MRI reconstruction. Using the ReconResNet model as the network backbone, the NCC1701 pipeline has been shown to be able to remove artefacts from highly undersampled images \citep{chatterjee2022reconresnet} with acceleration factors as high as 20. However, this work not only focuses on the magnitude images; it also ignores the phase images, which are fundamental requirements for MR thermometry \citep{chatterjee2022reconresnet}.

\subsection{Thermal Therapy and Thermometry}
Magnetic Resonance Imaging (MRI) has the ability to map temperatures \citep{cline1994mr,parker1983temperature}, and it has been more than 30 years since several extensive studies have been conducted to understand the quality of temperature monitoring in thermal treatment \citep{rossmann2014review}.
MR imaging provides a powerful, non-invasive tool for real-time temperature monitoring during minimally invasive thermal therapies. By utilising temperature-sensitive MRI parameters, clinicians can accurately measure and control temperature distributions within tissues, ensuring effective treatment while minimising damage to surrounding healthy tissue \citep{rieke2008mr}. The hotspot must be located correctly during ablation therapy with the use of MR guidance. It is necessary to locate the ablation site extremely precisely in order to burn only the unhealthy cells and spare the normal ones. Temperatures are achieved using microwave (MW), radio frequency (RF), ultrasound (US), or infrared (IR) techniques.
Thermal therapy can be divided into two techniques. Low temperature or hyperthermia (HT), where tumour tissue is heated to a temperature between 40 and 44 $^\circ C$  for 60 minutes with the aim of directly killing cancer cells, increasing oxygenation, and thus also increasing the radiosensitisation of the cancer cell \citep{kim1979clinical}. Local, regional, and whole body hyperthermia can be classified on the basis of the size of the heated area. External heat sources, as well as intraluminal or interstitial insertion of microwave-guided wires, can be used to apply heat to the tumour.
High-temperature thermal ablation, in which tumour tissues are heated to temperatures of 50-80 $^\circ C$ or higher for a shorter period of time, aims to kill cancer cells directly \citep{thomsen1991pathologic}.

\subsection{Deep learning in medical imaging}
The use of deep learning in the medical field, especially in the field of medical imaging, is increasing rapidly. Deep learning has achieved outstanding performance in the task of undersampled MR image reconstruction and the elimination of artefacts present in these MRIs \citep{qin2018convolutional, lyu2021cine}. \citet{wang2016accelerating} applied deep learning to compressed sensing MRI. 
Deep Residual Network (ResNet) was proposed by \citet{he2016deep} to optimise and improve the accuracy of deep learning models. ResNet was able to solve the vanishing gradient problem and open up the door to a deeper network. ResNet was proposed mainly for image classification, but it has later been used for many other applications, such as image classification \citep{mou2017unsupervised,zhang2019attention}, image segmentation \citep{pakhomov2019deep}, and image denoising \citep{jifara2019medical, zhu2017unpaired}. The residual learning model also proved to be very efficient in MRI reconstruction \citep{chatterjee2022reconresnet}.

One of the most commonly used network architectures for MRI reconstruction is UNet \citep{hyun2018deep}, which was first employed for the task of MR reconstruction in 2018. UNet is capable of reconstructing highly undersampled images.
\citet{chatterjee2022reconresnet} came up with the NCC1701 pipeline with the ReconResNet model as the backbone. This has shown an improvement in the reconstruction of undersampled Cartesian and radial MRIs over UNet, and it has demonstrated that it is capable of reconstructing up to acceleration factors of 20 and 17, for Cartesian and radial MRIs, respectively. 

UNet and ReconResNet work only in the image space (magnitude images), completely discarding the phase images. Although these methods work with both the image and the k-space, they apply real-valued convolution operations to complex-valued image space and k-space data, disrupting the rich geometric relationships within the complex data. In 2021, the first time complex-valued convolutions were applied for the task of undersampled MRI reconstruction directly in the k-space~\citep{chatterjee2021going}. 

On the other hand, \citet{adler2018learned} proposed the primal-dual network or PDNet, for the reconstruction of sparse computed tomography (CT) data. Given that CT and radial MRI reconstructions have mathematical similarities due to the Fourier slice theorem, \citet{ernst2021sinogram} applied PDNet successfully for the task of undersampled radial MRI reconstruction, and also extended PDNet to PDUNet, which outperformed PDNet with statistical significance. Both networks employ two types of network blocks, filtering in the image space and sinogram space. In 2022, these models were further extended using complex-valued convolutions into Fourier-PDNet and Fourier-PDUNet by~\citet{chatterjee2022complex}. These two models work in the k-space (i.e. the raw data space of MRI) instead of the sinogram space (i.e. the raw data space in the context of computed tomography), in addition to working in the image space. 

Although several models have been proposed for reconstructing undersampled MRIs, including models that work in both image and k-space, and models that work directly with complex data, the main focus of the evaluations carried out was on magnitude images. Current research aims to bridge this gap by focusing on both magnitude and phase images and then further evaluating the quality of MR thermometry from the reconstructed MRIs.

\subsection{MR-guided thermometry}

\begin{figure*}
\centering
\includegraphics[width=0.9\textwidth]{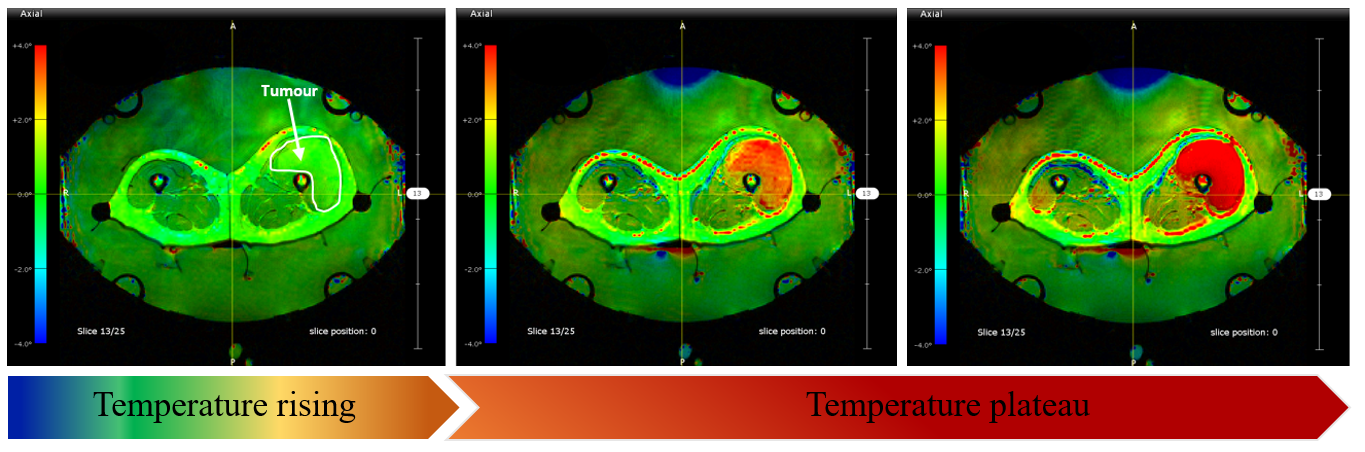}
\caption{An example of non-invasive PRF shift MR thermometry to monitor and control temperature during clinical hyperthermia (HT). The initial temperature map at time step t0 requires two MR images, which are calculated by voxel-wise subtracting the second phase image from the first reference phase image. This reference image is subtracted from the phase images of further acquired MR images, which are taken every 10 minutes during HT therapy. In the initial high-precision magnitude MR image, the temperature map is shown as a colour overlay (blue: relative temperature decrease; green: constant temperature; red: relative temperature rise).} \label{fig:bellini}
\end{figure*}

With the benefit of obtaining 3D temperature maps, MR-guided hyperthermia provides a non-invasive approach for temperature monitoring \citep{wust2006thermal}.On the basis of proton density, T1 or T2 relaxation time, the water molecule's molecular diffusion coefficient, magnetisation transfer, temperature-sensitive contrast agents, proton resonance frequency (PRF) shift imaging, or spectroscopy, various methods of measuring temperature with an MR system have been reported \citep{kuroda2005non, ludemann2010non, quesson2000magnetic, rieke2008journal, wlodarczyk1999comparison}. Techniques such as measuring longitudinal and transverse relaxation times \citep{parker1983temperature}, the diffusion coefficient, or the proton density rely heavily on the characteristics of the tissue. PRF shift imaging is independent of the tissue type and provides good linearity, and a desent temperature sensitivity. Because of this, the PRF shift technique is now the preferred technique for MRI-based temperature measurements due to its potential for online imaging and tumour control during treatments \citep{odeen2019magnetic,cernicanu2008validation, gellermann2005methods}. The PRF shift method's pre-clinical calibrations and uses are outlined in \citet{mcdannold2005quantitative}. The current standard for non-invasive temperature assessments in daily clinical practice is the PRF shift measurement.
The PRF-based phase mapping method stands out due to its linearity and reliability across different tissue types. Advances in MR imaging techniques continue to improve the precision and efficacy of thermal therapies \citep{rieke2008mr}.
The goal of the guided system is to understand the real-time temperature distribution and deliver quality controlled treatment and also be able to co-relate treatment temperature with treatment outcome in terms of actual thermal tissue damage. Fig. \ref{fig:bellini} shows an example of MR-based temperature monitoring at different time points.

\subsection{Contributions}
This research introduces Fourier-PDNet and Fourier-PDUNet – complex-valued neural networks that reconstruct undersampled MRIs, preserving the rich geometric structure of complex MRI data. As the MRI data are obtained in Fourier space – a complex data space – and the reconstructed images are also complex-valued, complex-valued convolutions would be essential to preserve the structure of the data properly. These methods are employed and evaluated here for the task of reconstructing highly undersampled (up to an acceleration factor of 10) MR thermometry data in terms of both reconstruction quality and the quality of the temperature maps obtained afterwards. To the authors' best knowledge, this is the first research addressing the problem of undersampled MR thermometry data (including hyperthermia). Moreover, the methods proposed here can also be used for the reconstruction of other types of undersampled MRI (including undersampled dynamic MRI). 

\section{Methodology}

\noindent Most of the previous work took either of these two directions: working only with the magnitude images (ignoring the phase images completely) or working with the complex image by splitting the data into real and imaginary parts before supplying it to the network as two separate channels. The first approach is not suitable for the current task at hand, while the second approach destroys the rich geometric structure present in the complex data. Both of these approaches apply real-valued convolution operations. As the data are complex-valued, applying complex-valued convolution should be better suited, which is capable of working directly with the complex-valued data without splitting them into channels, effectively preserving the geometric structure. 

\subsection{Experiment Design}
\begin{figure*}
\centering
\includegraphics[width=0.9\textwidth]{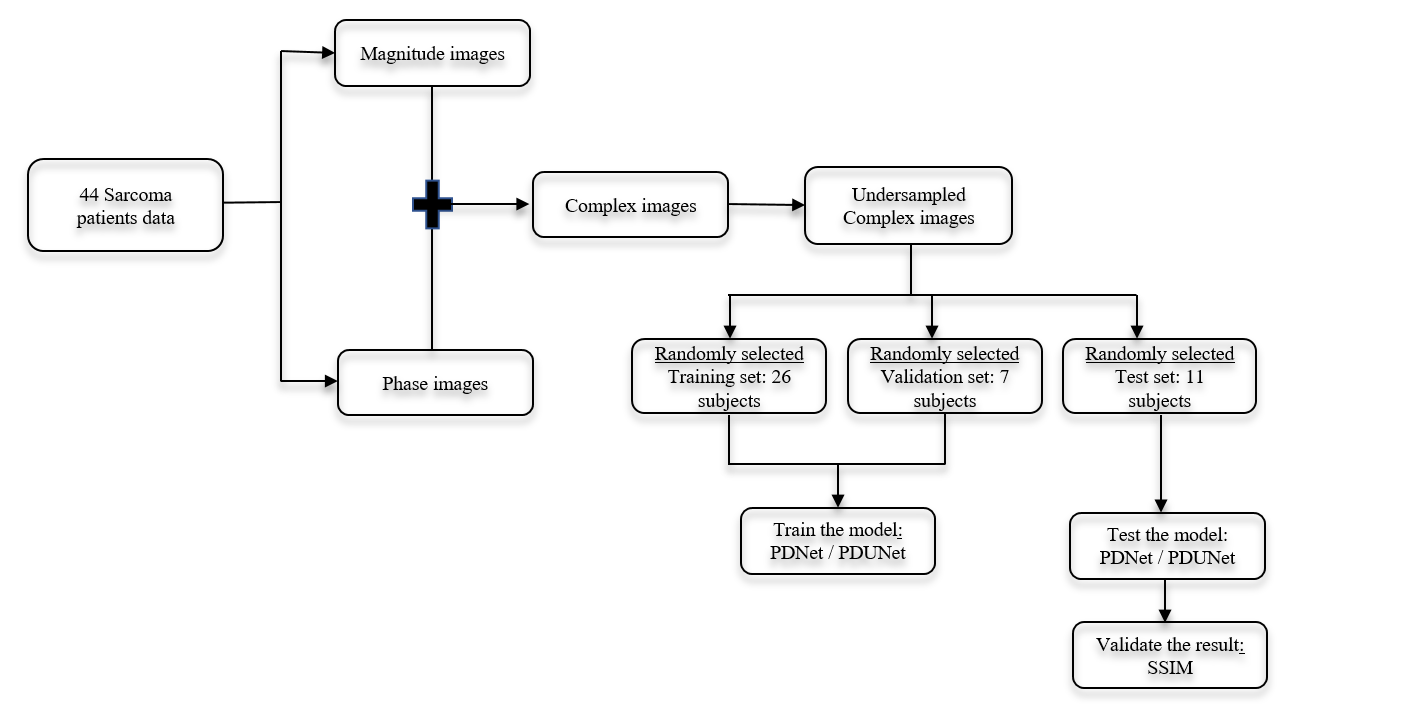}
\caption{Experiment Design} \label{fig:dataflow}
\end{figure*}

 MR images of 44 patients with different sarcoma cancers who have received the HT treatment in a combination of radiation/chemotherapy were used in this study (refer to Fig. \ref{fig:dataflow}). All methods were carried out in this study are in accordance with the ethical standards and approval (Application no. 24-168-Br) of the institutional research ethics committee of the Universit\"atsklinikum Erlangen, Friedrich-Alexnder-Universit\"at Erlangen-N\"urnberg, Erlangen, Germany, and with the 1964 Declaration of Helsinki and its later amendments. For this retrospective study, the requirement for formal consent was waived based on local legislation (BayKrG Art. 27 (4)).  
 
 One key goal of this work is the reconstruction of the temperature, so both magnitude and phase images are necessary. As the next step, the magnitude and phase images were combined together, and complex images were created. These complex images are artificially undersampled. Afterwards, the undersampled complex images were randomly divided into three different sets - Training, Validation and Test sets and the number of subjects was 26, 7 and 11, respectively. In the following steps, training and validation sets were used to learn the model weights of the modified PD Net \citep{adler2018learned} / PDUNet \citep{ernst2021sinogram} models, and the test set was used to evaluate the final performance. After testing, to quantitatively evaluate the results produced by the models, the Structural Similarity Index (SSIM) \citep{renieblas2017structural}) has been used. 

\subsection{Network Architectures}
Primal-Dual network or PD Net: The Primal-Dual network is a deep learning-based technique for computed tomography data with sparse sampling \citep{adler2018learned}. 
The algorithm unrolls a proximal-dual method with convolutional neural networks in place of the proximal operators to accommodate for (potentially non-linear) forward operators in deep neural networks. The algorithm is trained end-to-end, using only the raw measured data, and is not dependent on any initial reconstruction, such as filtered backprojection. This not only raises the standard of the final reconstruction, but also ensures data consistency. The quality of PD Net depends on the number of iterations, just like many iterative algorithms, for coverage of all the parameters of the network, an optimal number of iterations is needed. The fewer the parameters of the convolutional block, the more iterations are needed for convergence.

Primal-Dual UNet or PD UNet: Primal-Dual UNet\citep{ernst2021sinogram} is the improved version of the Primal-Dual network in terms of accuracy and reconstruction speed. A UNet has been used in place of a convolutional block of PD net for image space to obtain a higher number of parameters with low processing time. 

\begin{figure*}
\centering
\includegraphics[width=0.9\textwidth]{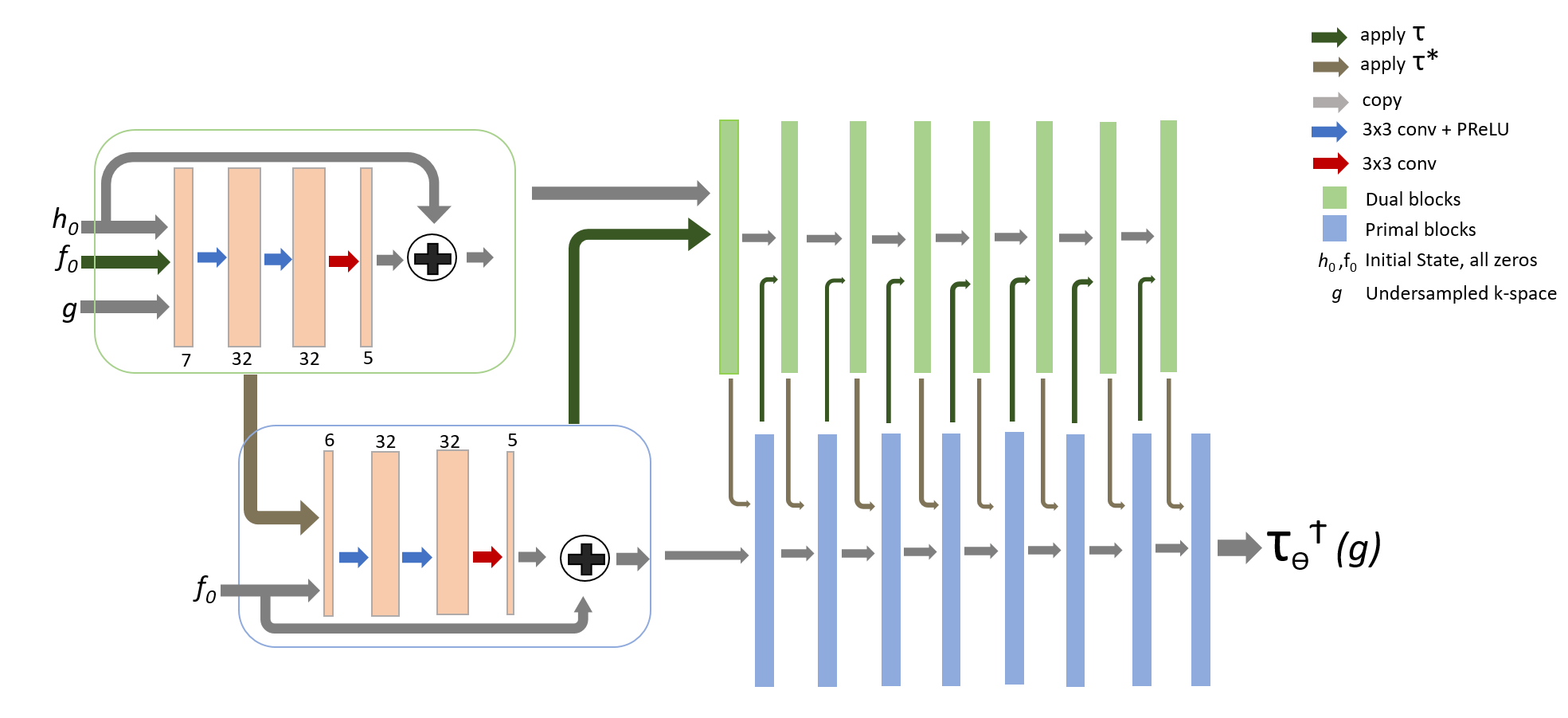}
\caption{ 
Fourier Primal-Dual Network (Fourier-PDNet) - modified version of the Primal-Dual network employing complex-valued convolution operations. Primal iterates are displayed in blue boxes, whereas dual iterates are displayed in green boxes. The architecture of all the blue boxes is the same and is shown in the matching large boxes. When several arrows lead to the same block, concatenation occurs before supplying the input to the first layer of the block. As the data are transmitted to the dual iterates, the initial estimates enter from the left. The primal blocks are responsible for removing artefacts from the image, while dual blocks attempt to predict the missing k-space frequencies. 
} 
\label{fig:PDNet}
\end{figure*}

\begin{figure*}
\centering
\includegraphics[width=0.9\textwidth]{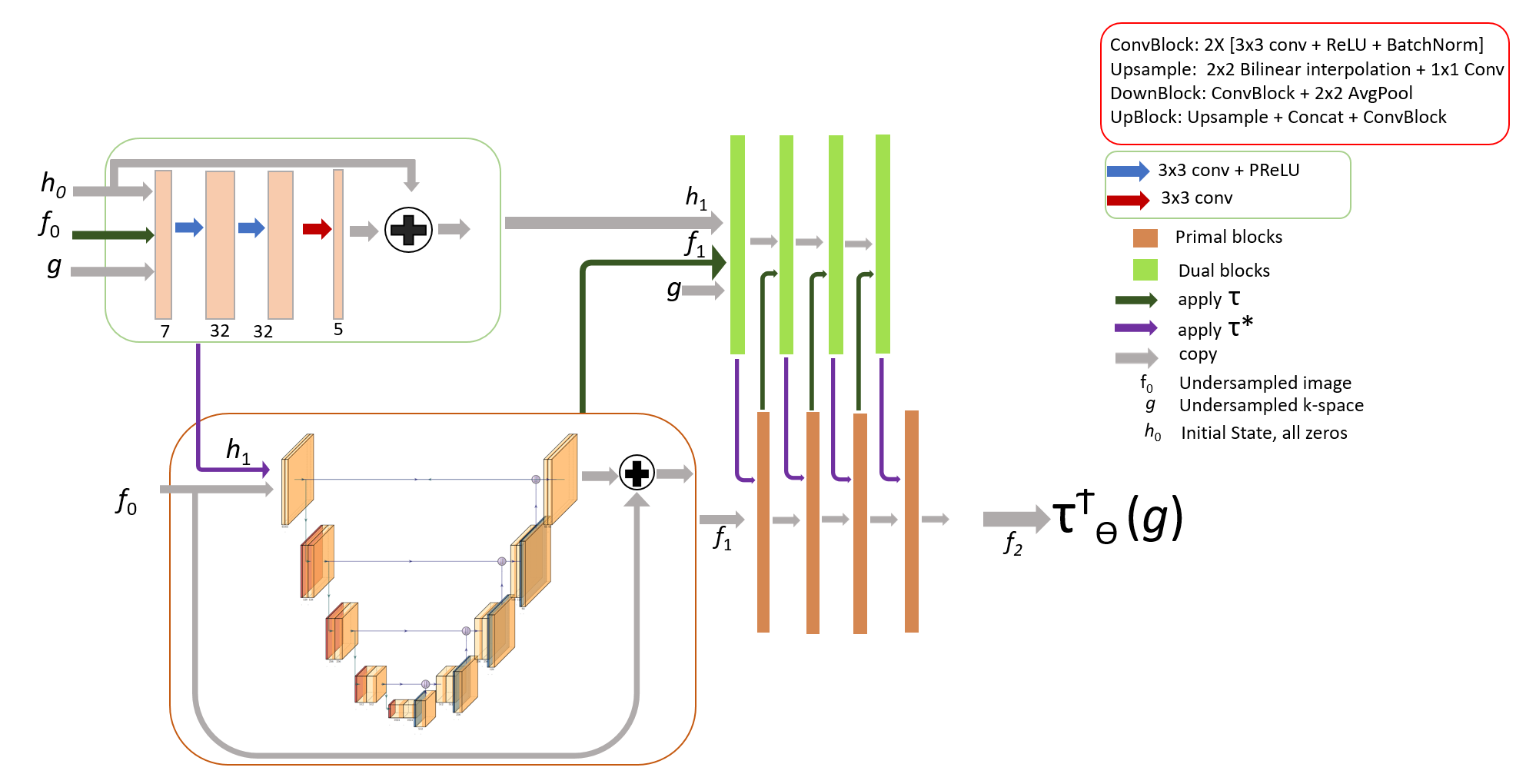}
\caption{Fourier Primal-Dual UNet (Fourier-PDUNet): The primal and dual iterates are represented by orange and green boxes, respectively. A complex-valued UNet architecture, as opposed to a complex-valued fully convolutional network (as used in Fourier-PDNet), is used in the primal block. The original primal-dual network is still present in the dual block, which is a complex-valued fully convolutional network. 
} \label{fig:PDUNet}
\end{figure*}

In this study, these two network models were modified by employing complex-valued convolutions (see Sec. \ref{sec:implementation}) to be able to work with complex-valued data, resulting in Fourier-PDNet and Fourier-PDUNet models (shown in Figures \ref{fig:PDNet} and \ref{fig:PDUNet}, respectively). 


\subsection{Data consistency step}
 In the data consistency step, the actual acquired undersampled data replaces the network's output. The network only helps to fill the data which were ignored before during the undersampled data, this is how the final output is not totally dependent on the network.
 
 Following \citep{hyun2018deep}, a data consistency step was performed for the after reconstructing the undersampled Cartesian data. To obtain the corresponding k-space, FFT was performed on the output image first. Then, to identify the k-space values that were not acquired, an inverted mask was applied to this. The missing estimated k-space data from the network were combined with the measured data. To obtain the final output, iFFT was applied to this combined k-space.

\subsection{Dataset}
In this work, MRIs of 44 patients treated in the Department of Radiation Oncology of the Universitätsklinikum Erlangen, acquired from 2015 to 2020, who underwent HT treatment with MR thermometry, were used. The image sets of 44 different patients with different types of sarcoma cancer, originating mainly in the leg of the patients (details are in Table \ref{tab:Dataset2}), have been acquired at Siemens Magnetom Symphony 1.5T scanner (Siemens Healthineers AG, Erlangen, Germany), the scanning sequence is GR or Gradient Recalled, Sequence Name fl2D or Fast low angle shot (FLASH 2D). A total of 24,486 MRI 2D slices, across 138 of treatment sessions, have been utilised in this work. Each subject’s static and dynamic scans were acquired in different sessions using the same sequence and parameters. The age range of the patients is 23 to 80 years. 26 subjects out of 44 were used for the training set selected randomly; seven subjects were used for the validation set, and 11 subjects were used for testing the model. MR thermometry images have two types of acquisition, static acquisition and dynamic acquisition, and the parameters of these two types of acquisition have been shown in Table \ref{tab:Static MR acquisition information} and in Table \ref{tab:Dynamic MR acquisition information}, respectively.

\begin{table}[htbp]
\caption{Static MR acquisition information}\label{tab:Static MR acquisition information}
\begin{center}
\resizebox{0.8\columnwidth}{!}{%
\begin{tabular}{@{}cc@{}}
\toprule
\textbf{Parameter}     & \textbf{Value}            \\ \midrule
Field strength         & 1.5 T                     \\
Scanning Sequence      & GR                        \\
Sequence Name          & fl2D                      \\
Acquisition Type       & 2D                        \\
Repetition time (TR)   & 120 ms                    \\
Echo time (TE)         & 4.76ms \& 19.1ms           \\
Flip angle             & 70$^\circ$             \\
Bandwidth             & 260 Hz/Px            \\
Voxel size             & $1.95x1.95x(1 to 13)mm^3$ \\ 
Acquisition duration    & 33 to 93 secs \\ \bottomrule
\end{tabular}%
}
\end{center}
\end{table}

\begin{table}[htbp]
\caption{Dynamic MR acquisition information}\label{tab:Dynamic MR acquisition information}
\begin{center}
\resizebox{0.8\columnwidth}{!}{%
\begin{tabular}{@{}cc@{}}
\toprule
\textbf{Parameter}     & \textbf{Value}                      \\ \midrule
Field strength         & 1.5 T                               \\
Scanning Sequence      & GR                                  \\
Sequence Name          & fl2D                                \\
Acquisition Type       & 2D                                  \\
Repetition time (TR)   & 600 ms                              \\
Echo time (TE)         & 4.76ms \& 19.1ms                    \\
Flip angle             & 50$^\circ$                  \\
Bandwidth             & 150 or 260 Hz/Px            \\
Voxel size             & $3.90\times 3.90\times(1 to 5)mm^3$ \\  
Acquisition duration    & 118 to 225 secs \\ 
Number of TPs    & 2 to 17 \\
Time per TP    & 12 to 89 secs \\\bottomrule
\end{tabular}%
}
\end{center}
\end{table}

\begin{table}[htbp]
\caption{Data set}\label{tab:Dataset2}
\begin{center}
\resizebox{0.8\columnwidth}{!}{%
\begin{tabular}{@{}cc@{}}
\toprule
\textbf{Type of cancer}    & \textbf{Number of Patients} \\ \midrule
Liposarcoma                & 15                          \\
Pleomorphic sarcoma        & 10                          \\
Synovial Sarcoma           & 7                           \\
Leiomyo sarcoma            & 5                           \\
Soft tissue sarcoma        & 3                           \\
Rhabdomyo sarcoma          & 2                           \\
Spindle cell sarcoma       & 2                           \\
Myxofibro sarcoma          & 1                           \\
Pleomorphic LeiomyoSarcoma & 1                           \\
Ewing sarcoma              & 1                           \\
Fibro sarcoma              & 1                           \\ \bottomrule
\end{tabular}%
}
\end{center}
\end{table}that several extensive studies have been performed

\subsection{Undersampling}
For the Cartesian sampled experiment, all images from different subjects were treated as fully sampled images. As the datasets do not contain any raw MR data, using the MR-Under \citep{chatterjee2020soumickmj} pipeline, the single-channel fully sampled raw data and various undersampled datasets were generated artificially.

\begin{figure}
\centering
\includegraphics[width=0.49\textwidth]{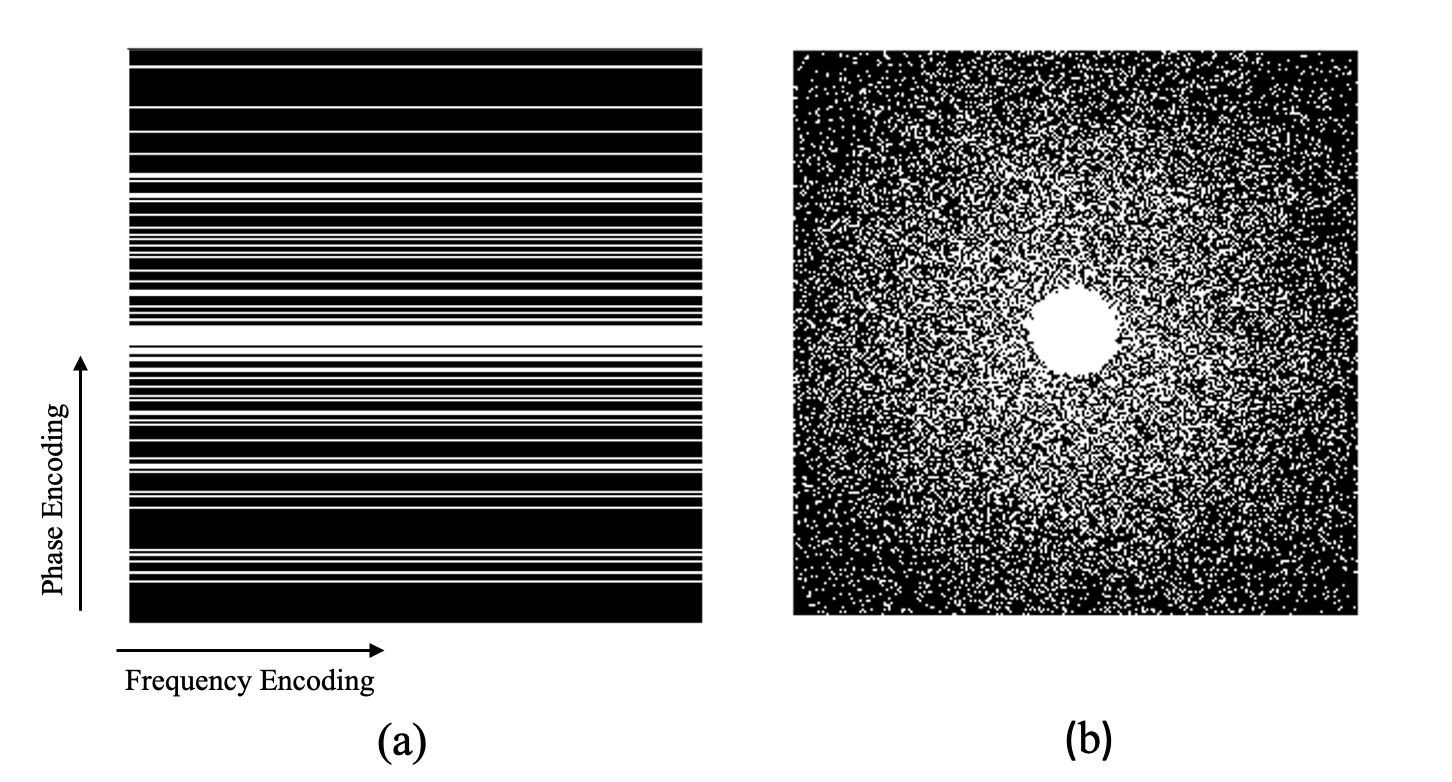}
\caption{(a) 1D Varden mask and
(b) 2D Varden mask. All of them are for image size 256x256, taking 25\% of
the k-space.} \label{fig:vardenmask}
\end{figure}

 Cartesian raw data have been artificially undersampled using the k-space sampling pattern \citep{lustig2007sparse}, also known as the sampling mask, which was created by randomly choosing completely sampled readout lines in the phase encoding direction, with the centre of the distribution following a one-dimensional normal distribution (Fig. \ref{fig:vardenmask}(a)) that matches the k-centre space (referred to as 1D Varden). This sampling mask consisted of a densely sampled centre consisting of eight lines, while gradually decreasing the sampling density toward the edges of the k-space \citep{lustig2007sparse}. Furthermore, another mask is designed (Fig. \ref{fig:vardenmask}(b)), referred here as 2D Varden, that contains a densely sampled centre covering 2.5\% of the k space, while the rest of the k-space is sampled randomly and distributed according to a two-dimensional normal distribution pattern \citep{lustig2007sparse}. Three distinct Cartesian undersampling patterns were used in the first round of trials. 1D and 2D Varden masks were generated by randomly sampling 25\% or 10\% of the k-space, achieving acceleration factors of 4 and 10, respectively. 

\subsection{Implementation}
\label{sec:implementation}
As input, the complex undersampled images are used to train the network to obtain the reconstructed complex images. However, most deep learning networks are implemented only on real-valued data, not complex-valued data. Therefore, the use of complex-valued convolution or CV-CNN\citep{chatterjee2022complex}  was a necessity. 
The convolution operation is the main component of CNNs, and it is computed by the sum of the product of two functions - the input (x) and the kernel (w) and the outcome is referred to as the feature map or activation map (s), and it is given by:
\begin{equation}
    s(t)=(w \star x)(t)=\sum_a x(t+a) w(a)
\end{equation}
In this case, w and x are both real-valued. Complex-valued convolutional networks, commonly referred to as CV-CNNs, improve on this by using the complex-valued convolution operation, which is defined as:
\begin{equation}
\begin{aligned}
C_r(t) & =\left(w_r \star x_r\right)(t)-\left(w_i \star x_i\right)(t) \\
C_i(t) & =\left(w_i \star x_r\right)(t)+\left(w_r \star x_i\right)(t)
\end{aligned}
\end{equation}
where $x_r$ and $x_i$ are the real and imaginary components of the complex-valued input $x$, respectively. Similarly, $w_r$ and $w_i$ are components of the complex-valued kernel $w$, and $C_r$ and $C_i$ are components of the generated complex-valued feature map s. This can also be expressed in matrix notation:
\begin{equation}
\left[\begin{array}{c}
\Re(\mathbf{w} \star \mathbf{x}) \\
\Im(\mathbf{w} \star \mathbf{x})
\end{array}\right]=\left[\begin{array}{cc}
\mathbf{w}_{\mathbf{r}} & -\mathbf{w}_{\mathbf{i}} \\
\mathbf{w}_{\mathbf{i}} & \mathbf{w}_{\mathbf{r}}
\end{array}\right] \star\left[\begin{array}{c}
\mathbf{x}_{\mathbf{r}} \\
\mathbf{x}_{\mathbf{i}}
\end{array}\right]  
\end{equation}
CV-CNNs can learn more sophisticated representations while preserving the algebraic structure of complex-valued data.
Fig.\ref{fig:WorkingMethod} shows the working mechanism of the proposed framework.

\subsection{Training and Inference}
\begin{figure*}
\centering
\includegraphics[width=0.9\textwidth]{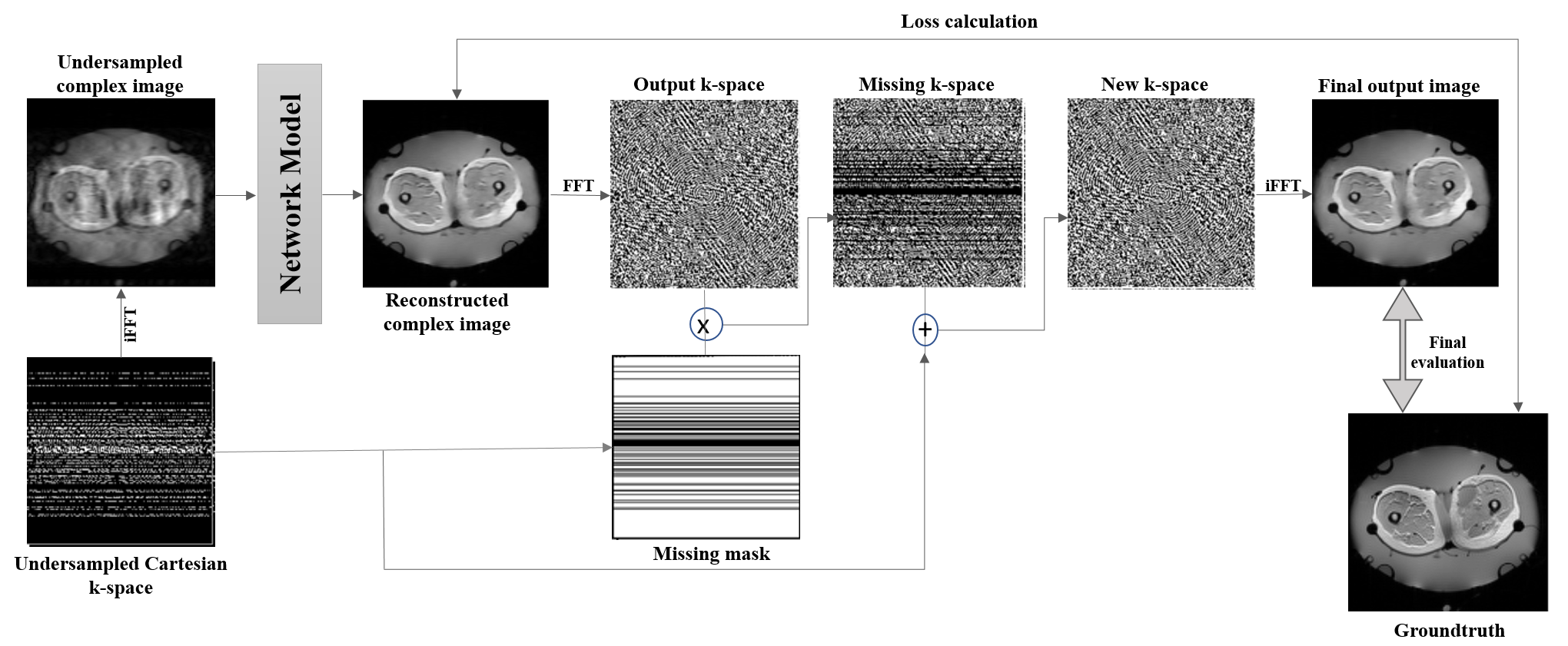}
\caption{Workflow of Neural Network-Based MRI Reconstruction from Undersampled k-Space Data: This figure outlines the workflow for reconstructing high-quality MRIs from undersampled k-space data using a neural network. Starting with undersampled Cartesian k-space data, an Inverse Fourier Transform (iFFT) generates an artefact-laden complex image. This image is fed into a neural network to produce a reconstructed complex image. The network's output undergoes a Fast Fourier Transform (FFT) to form output k-space, which is combined with the original sampled k-space data using a missing mask. A final Inverse Fourier Transform (iFFT) converts the combined k-space back to the spatial domain, resulting in the final output image. The image is then compared to the ground truth for loss calculation and evaluation, guiding the network to improve its reconstruction accuracy.} \label{fig:WorkingMethod}
\end{figure*}

Fig.\ref{fig:WorkingMethod} demonstrates the operational principles of the complete framework, which include a network backbone and a data consistency step. During the training process, only the network backbone was used. However, the entire framework is used during inference. This framework is identical to the NCC1701 framework \citep{chatterjee2022reconresnet}, except for the backbone model. The original ReconResNet backbone was replaced with the complex-valued models Fourier-PDNet and Fourier-PDUNet to be able to reconstruct both magnitude and phase images. The loss function to train the backbone model of the original NCC1701 was also replaced with a complex-valued version of L1 loss or mean absolute error (MAE). The L1 loss between a prediction and the actual value is calculated using:

\begin{equation}
\begin{aligned}
   f(y, \hat{y}) = \frac{1}{N} 
   \sum_{n=1}^N \left( \left| y_{r,n} - \hat{y}_{r,n} \right| + \left| y_{i,n} - \hat{y}_{i,n} \right| \right)
\end{aligned}
\label{eq:MAE}
\end{equation}
Where $y$ is the actual value or ground truth, $\hat{y}$ is the predicted value, and $N$ is the number of samples in the whole dataset. Here, both $y$ and $\hat{y}$ are complex-valued with real $r$ and imaginary $i$ parts.

The model was trained for 100 epochs, with a batch size of one, and the loss value was then minimised using Adam Optimiser (Initial learning rate 0.0001, decayed by 10 after every 50 epochs; $\beta_1 = 0.9, \beta_2 = 0.999, \epsilon = 1e-09$).
This network was implemented using PyTorch~\citep{paszke2019pytorch}, Python version 3.10.9 was used and was trained using NVIDIA GeForce RTX 2080 Ti.

\subsection{Evaluation Criteria}
Structural Similarity Index \ref{eq:SSIM}(SSIM) \citep{renieblas2017structural}, Normalised root-mean-squared error (NRMSE, \ref{eqn:NRMSE}) and Universal Image Quality Index (UIQI) \citep{wang2002universal} have been used to evaluate the results.

The range of SSIM values is between zero and one, where the higher the SSIM value, the higher the similarity between two images. 

\begin{equation}
    SSIM (x,y) = \frac{(2\mu_x\mu_y+C_1)(2\sigma_xy+C_2)}{(\mu_x^2+\mu_y^2+C_1)(\sigma_x^2+\sigma_y^2+C_2)}
    \label{eq:SSIM} 
\end{equation}
where $x$ and $y$ are the two images between which the structural similarity is to be calculated, $\mu_x, \mu_y, \sigma_x, \sigma_y$ and $\sigma_{xy}$, are the pixel means of $x$, pixel means of $y$, standard deviations, and cross-covariance for images x and y, respectively. $c_{1}=(k_{1}L)^{2}$ and $c_{2}=(k_{2}L)^{2}$ where $L$ is the dynamic range of the pixel-values, $k_{1}=0.01$ and $k_{2}=0.03$.

To statistically compare the two images (output and ground truth), NRMSE was used, calculated as:

\begin{equation}
\begin{aligned}
    {MSE} ={\frac {1}{n}}\sum _{i=1}^{n}(Y_{i}-{\hat {Y_{i}}})^{2}
\end{aligned}
\label{eq:mse}
\end{equation}

\begin{equation}
N R M S E=\frac{\sqrt{M S E}}{\sqrt{\frac{1}{n} \sum_{i=1}^{n} Y^2}}
\label{eqn:NRMSE}
\end{equation}

where the pixels of the fully sampled ground truth image have been denoted as $Y_{i}$, the pixels of the undersampled image or the reconstruction (depending on the comparison performed) have been denoted as $\hat{Y_{i}}$ and $n$ denotes the number of pixels in the image.

Universal Image Quality Index (UIQI)\citep{wang2002universal} Loss of correlation, luminance distortion, and contrast distortion are the three components that make up any image distortion when it is modelled. The proposed index is simple to calculate and adaptable to numerous image processing applications, as opposed to using conventional error summation techniques.

\begin{equation}
    Q = \frac{\sigma_{x_y}}{\sigma_x \sigma_y} \cdot \frac{2\Bar{x}\Bar{y}}{(\Bar{x})^2 + (\Bar{y})^2} \cdot \frac{2\sigma_x\sigma_y}{\sigma^2_x + \sigma^2_y}
    \label{eq:UIQI} 
\end{equation}
\[
\begin{gathered}
\bar{f}=\frac{1}{M N} \sum_{i=0}^{M-1} \sum_{j=0}^{N-1} x[i, j] \quad \bar{y}=\frac{1}{M N} \sum_{i=0}^{M-1} \sum_{j=0}^{N-1} y[i, j] \\
\sigma_{x y}=\frac{1}{M+N-1} \sum_{i=0}^{M-1} \sum_{j=0}^{N-1}(x[i, j]-\bar{x})(y[i, j]-\bar{y}) \\
\sigma_x^2=\frac{1}{M+N-1} \sum_{i=0}^{M-1} \sum_{j=0}^{N-1}(x[i, j]-\bar{x})^2 \\
\sigma_y^2=\frac{1}{M+N-1} \sum_{i=0}^{M-1} \sum_{j=0}^{N-1}(y[i, j]-\bar{y})^2
\end{gathered}
\]
where x and y are two images, considered as matrices having M and N number of columns and rows with x[i,j], y[i,j] pixels where (0$\geq$i $>$ M, 0 $\geq$j $>$ N ) and $Q $ is the Universal image quality index.

$Q$  can be obtained by multiplying three components together. The correlation coefficient is the initial component, which quantifies the level of linear correlation between the images x and y; the range varies [-1,1]. 
The second component assesses the similarity of mean luminance between images and has a value range of [0, 1].
With a range of [0, 1], the third component quantifies how closely the contrasts of the images match.

\subsection{Temperature Map}
Temperature maps were generated to evaluate the retrieved temperatures by the Fourier-PDUNet and Fourier-PDNet models, as well as from the undersampled inputs and ground-truth images for comparison. To create these temperature maps, the Proton Resonance Frequency Shift (PRFS) \citep{mcdannold2005quantitative} method has been used. For MRI-based temperature measurements, the PRF shift approach is currently the clinically recommended practice.

In Gradient Recalled Echo (GRE) images, the change of resonance frequency is expressed as phase change. The temperature difference can be derived by calculating the phase difference between a GRE image at a certain temperature and a reference temperature \citep{ishihara1995precise}. The linear relationship between the temperature difference and the phase change can be expressed as the following equation.
\begin{equation}
    \Delta T = \frac{\phi (T) - \phi (T{ref})}{\Upsilon\alpha \beta  TE}
\end{equation}
where $\Delta T$ = Temperature difference,  $\alpha$ = Temperature sensitivity of PRFS, $\Upsilon$ = Gyro-magnetic constant,
$\delta$ = Main Magnetic field strength and TE = Echo time.
According to Eq.\ref{eq:phasediff}, a complex calculation has been performed to construct the phase difference, which could avoid the phase wrapping problem during the heating cycle \citep{peters2000magnetic}. 

\begin{equation}
\label{eq:phasediff}
    \Delta\phi = \text{atan} \left(\frac{\mathrm{Re}(I_{\text{ref}}) \cdot \mathrm{Im}(I_H) - \mathrm{Im}(I_{\text{ref}}) \cdot \mathrm{Re}(I_H)}{\mathrm{Re}(I_{\text{ref}}) \cdot \mathrm{Re}(I_H) + \mathrm{Im}(I_{\text{ref}}) \cdot \mathrm{Im}(I_H)} \right)
\end{equation}

where Re and $I_m$ are the real and imaginary components of the heated ($I_H$) and reference ($I_{ref}$) images.

\subsubsection{Comparison of the temperature maps}
As the focus of this research is hyperthermia (or even MR thermometry in general), it is not sufficient to evaluate only the reconstruction quality of the magnitude and phase images. Rather, it is important to evaluate the reconstructed temperature maps. Given that this paper used real clinical data that are not free of noise, and it is difficult to find noise-free non-heated regions in the temperature maps, conventional techniques, such as the temperature-to-noise ratio (TNR) \citep{madore2011multipathway} cannot be applied reliably. Hence, the accuracy of the resultant temperature maps - the error in reconstructing the temperature maps from the undersampled volumes and the models' outputs, compared to the temperature maps obtained from the fully-sampled ground-truth volumes were calculated following the equation:

\begin{equation}
\label{eq:temperror}
    E_T(x,y) = \frac{100}{V \cdot \tau} \sum_{t=1}^{\tau} \sum_{v=1}^{V} \left( \Delta T_{\text{x}(v, t)} - \Delta T_{\text{y}(v, t)} \right)^2
\end{equation}

where $x$ is the fully-sampled data, $y$ is the undersampled or reconstructed data, $V$ is the total number of 3D voxels, $\tau$ is the total number of time points, and finally, $\Delta T_{\text{x}(v, t)}$ and $\Delta T_{\text{y}(v, t)}$ represent the temperature values at voxel $v$ and time point $t$ for $x$ and $y$. Two sets of $E_T(x,y)$ were computed in this research: considering the whole volume, and only considering the region of interest by segmenting the tumour region. These demonstrate the overall error in terms of the temperature with respect to the whole volume and the region of interest.

\section{Results}

\subsection{Baseline comparison}
Initially, the performance of these complex-valued models was benchmarked against one of the MR reconstruction models – ReconResNet \citep{chatterjee2022reconresnet}. There were three main reasons behind choosing this model – ReconResNet is a stand-alone model that works directly with coil-combined images and not an end-to-end MRI reconstruction framework, the original paper demonstrated that this model works well with both 1D and 2D Varden masks, and finally, because it is the originally proposed backbone model for the NCC1701 framework that was also used here. However, this model was originally proposed to reconstruct only the magnitude images, and phase image reconstruction is essential for MR thermometry. Complex data can be supplied to a real-valued model in two different ways: a 2-channel input containing real and imaginary parts, or magnitude and phase. The ReconResNet model was modified to take input and produce output with two channels. For the baseline purposes, both possibilities were evaluated - real+imaginary and magnitude+phase for reconstructing MRIs undersampled with 1D Varden 25\%.

\begin{figure*}
\centering
\includegraphics[width=\textwidth]{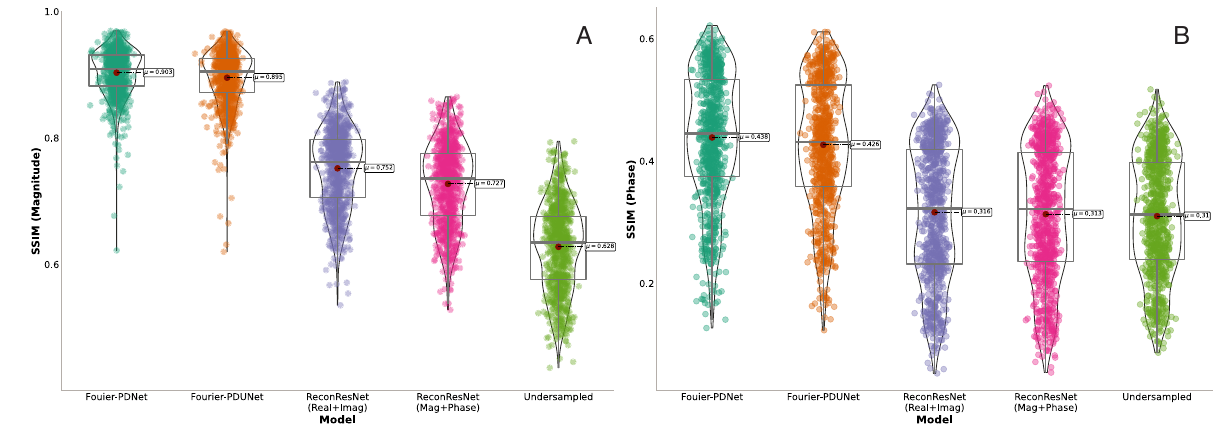}
\caption{Comparison of the reconstruction quality achieved by the complex-valued models—Fourier-PDNet and Fourier-PDUNet—and the real-valued model ReconResNet with two types of inputs—real and imaginary, and magnitude and phase—as well as the zero-padded k-space (denoted as undersampled) for k-space undersampled using 1D variable density sampling taking 25\% of the k-space. SSIM values for (A) magnitude and (B) phase.} \label{fig:baselines}
\end{figure*}

Fig. \ref{fig:baselines} demonstrates the results in terms of SSIM for (A) magnitude and (B) phase, obtained from the real-valued baseline models — ReconResNet (real + imaginary) and ReconResNet (magnitude + phase), and from the complex-valued models — Fourier-PDNet and Fourier-PDUNet. These results are then compared against the zero-filled k-space reconstruction (denoted as undersampled).

The real-valued baseline models ReconResNet (Real + Imag) and ReconResNet (Magnitude + Phase) resulted in 76\% and 74\% median SSIM scores for the magnitude images, while achieving 33\% and 32\% for the phase images, respectively. These scores improved upon the undersampled (zero-filled) reconstruction, which achieved 63\% and 31\%, respectively. The complex-valued models resulted in even higher scores than the real-valued baselines; Fourier-PDNet and Fourier-PDUNet achieved 91\% and 90\% SSIM for the magnitude images and 44\% and 40\% SSIM for the phase images, respectively. All improvements observed were statistically significant, as determined using the Wilcoxon signed-rank test. Hence, it can be concluded that the complex-valued models significantly outperformed the real-valued baseline models in reconstructing both magnitude and phase images, and all further in-depth analyses were performed using the complex-valued models only.

\subsection{Evaluation of the complex-valued models}

\begin{figure*}
\centering\includegraphics[width=0.8\textwidth]{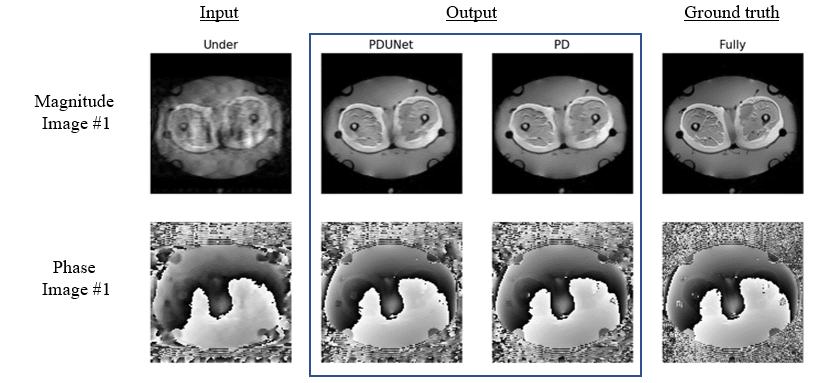}
\caption{Qualitative Result: 1D Varden 25\% - Comparison of MR Image Reconstruction Methods from Undersampled Data. This figure illustrates the qualitative results of MR image reconstruction from 1D Varden 25\% undersampled k-space data, comparing two methods: PDUNet and PD. The input column displays the undersampled images with visible artefacts. The PDUNet and PD columns show the reconstructed outputs, while the ground truth column provides the fully sampled reference images.}
\label{fig:QualitativeReslt1D25}
\end{figure*}

\begin{figure*}
\centering
\includegraphics[width=\textwidth]{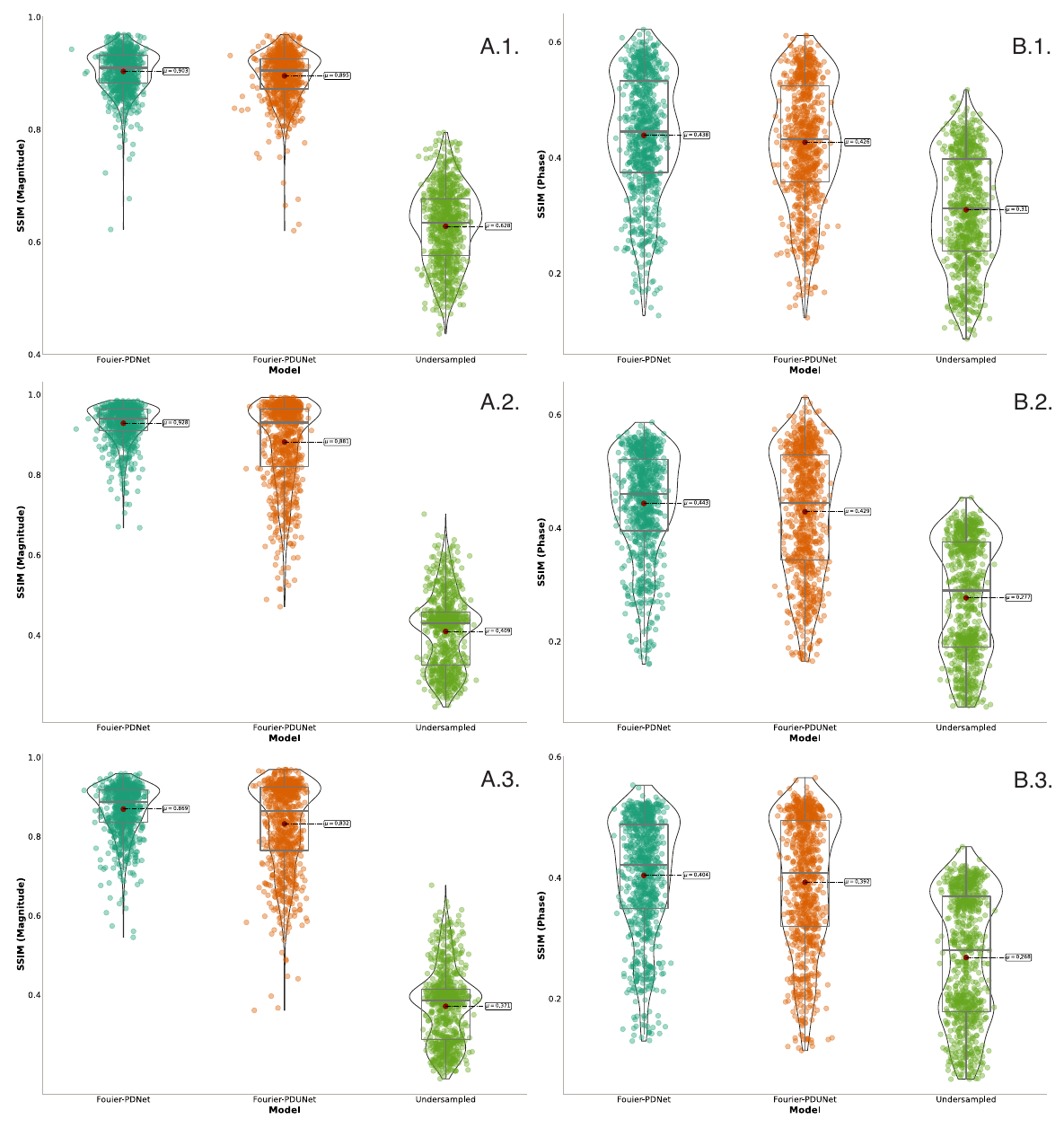}
\caption{The reconstruction quality achieved by the complex-valued models - Fourier-PDNet and Fourier-PDUNet, compared against the zero-padded k-space (denoted as undersampled), for k-space being undersampled using (x.1.) 1D variable density sampling taking 25\% of the k-space, (x.2.) 1D variable density sampling taking 25\% of the k-space, and (x.3.) 2D variable density sampling taking 10\% of the k-space. SSIM values for (A.n.) magnitude and (B.n.) phase are presented.} \label{fig:NewQualitativeReslt1}
\end{figure*}

\noindent Images from 44 patients with sarcoma cancer were undersampled with an acceleration factor of 4, resulting in average SSIM values of 1D varden 25\% is 63\% and 31\%, for magnitude and phase images respectively, where the Fourier-PDNet and Fourier-PDUNet models managed to reconstruct those data with average SSIM values of 91\% and 90\% for magnitude images, while achieving 44\% and 40\% for phase images, respectively. The results are displayed using the violin plots in Fig. \ref{fig:NewQualitativeReslt1}. Example outputs from two different subjects for 1D varden 25\% sampling patterns are shown in Fig. \ref{fig:QualitativeReslt1D25} for qualitative evaluation.

\begin{figure*}
\centering
\includegraphics[width=0.8\textwidth]{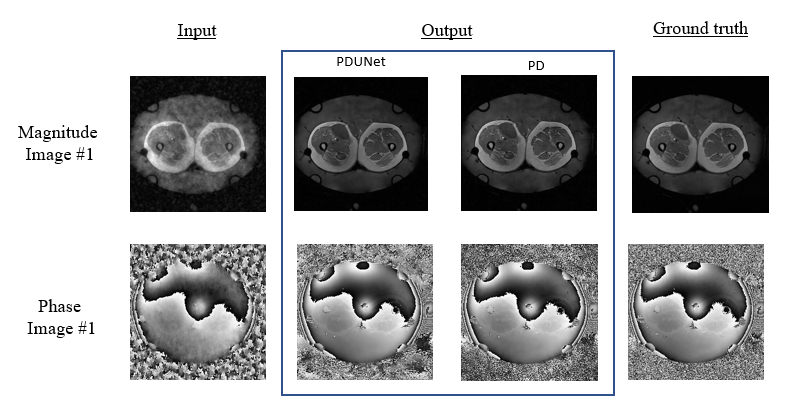}
\caption{Qualitative Result: 2D Varden 25\% - Comparison of MR Image Reconstruction Methods from Undersampled Data. This figure presents the qualitative results of MR image reconstruction from 2D Varden 25\% undersampled k-space data, comparing two methods: PDUNet and PD. The input column shows the undersampled images with noticeable artefacts. The PDUNet and PD columns display the reconstructed outputs, with PDUNet demonstrating superior artefact reduction and clearer images compared to PD. The ground truth column provides the fully sampled reference images. Both magnitude and phase images are included.}
\label{fig:QualitativeReslt2D25}
\end{figure*}

The average SSIM values of the 2D varden undersampled MRIs with 25\% undersampling were 43\% and 29\% for the magnitude and phase images. The Fourier-PDNet and Fourier-PDUNet models managed to reconstruct those with average SSIM values of 94\% and 93\% for the magnitude images while achieving 47\% and 46\% for the phase images, respectively. The SSIM values using violin plots and example results are shown in Figures \ref{fig:NewQualitativeReslt1} and \ref{fig:QualitativeReslt2D25}, respectively.

\begin{figure*}
\centering
\includegraphics[width=0.8\textwidth]{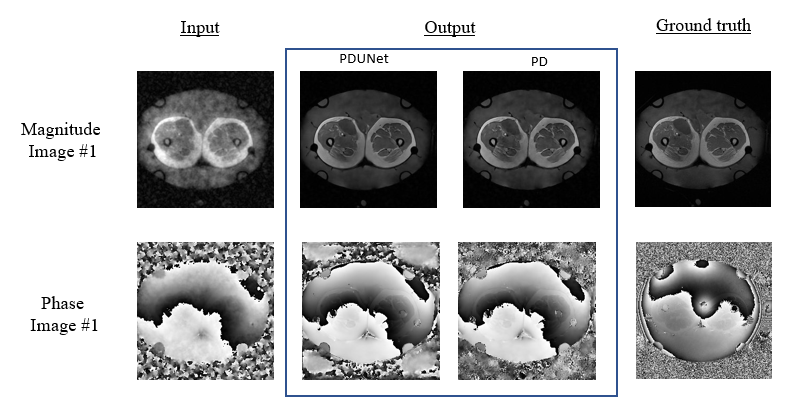}
\caption{Qualitative Result: 2D Varden 10\% - Comparison of MR Image Reconstruction Methods. This figure presents MR image reconstruction results from 2D Varden 10\% undersampled k-space data, comparing PDUNet and PD methods. The input column shows undersampled images with artefacts. The PDUNet and PD columns display reconstructed outputs, with PDUNet providing superior artefact reduction and clearer images. The ground truth column shows fully sampled reference images. Both magnitude and phase images are included.} \label{fig:QualitativeReslt2D10}
\end{figure*}

Finally, with an acceleration factor of 10, the SSIM values of the undersampled images of the 2D varden 10\% were 39\% and 28\% for the magnitude and phase images, respectively. Fourier-PDNet and Fourier-PDUNet models improved the SSIM values to 87\% and 86\% for the magnitude images while achieving 43\% and 41\% for the phase images, respectively. The violin plots of the SSIM values and the qualitative comparison for two subjects are shown in Figures \ref{fig:NewQualitativeReslt1} and \ref{fig:QualitativeReslt2D10}, respectively.

Table \ref{tab:ResultTable} provides a complete qualitative overview of the results using NRMSE and UIQI, along with the SSIMs. 

\begin{figure*}
\centering
\includegraphics[width=0.8\textwidth]{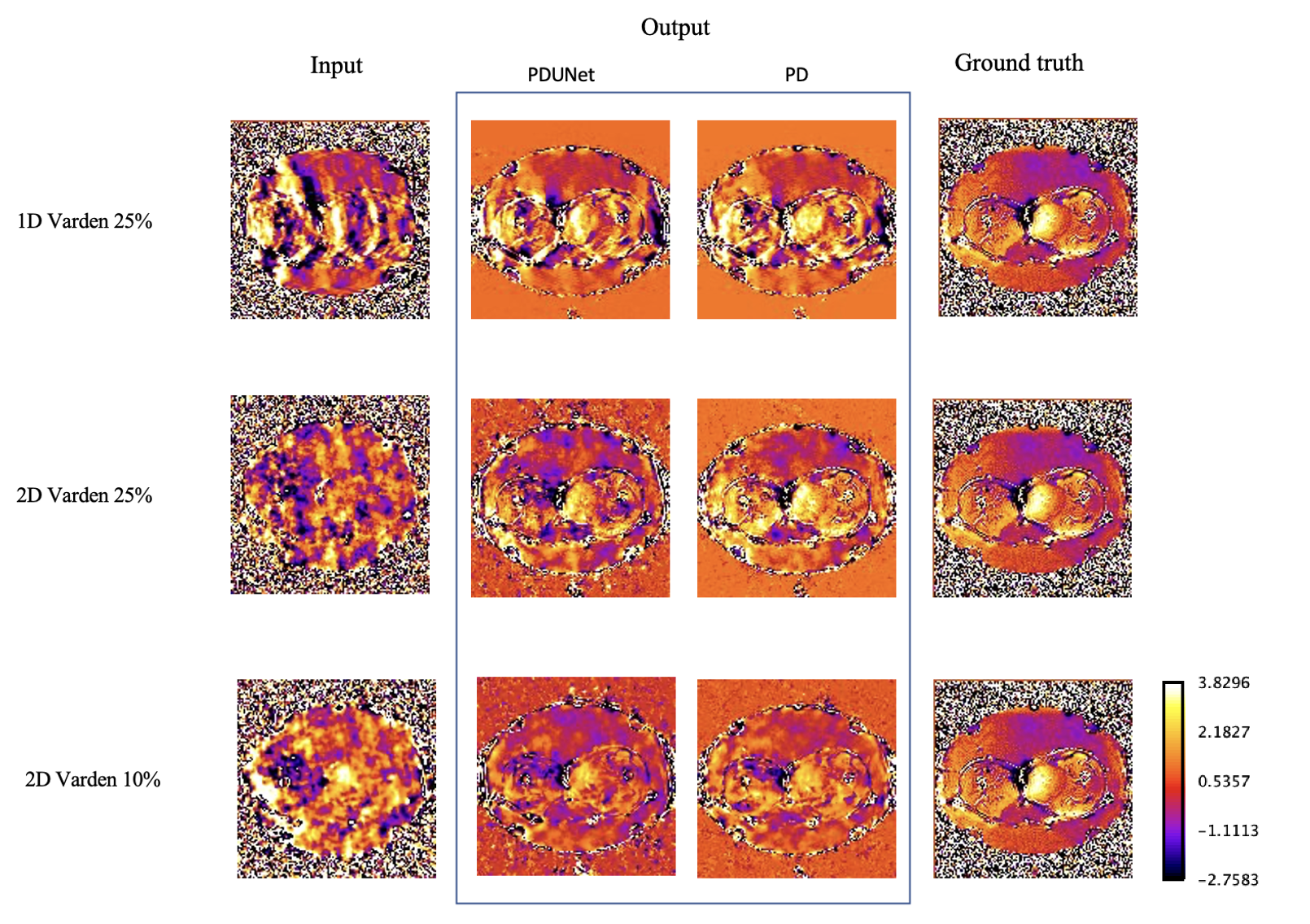}
\caption{Reconstructed Temperature Maps: This figure illustrates the qualitative results of MR image reconstruction from different undersampled k-space data sets, comparing two methods: PDUNet and PD, against the ground truth. The datasets include 1D Varden 25\%, 2D Varden 25\%, and 2D Varden 10\% undersampled k-space data. The maps demonstrate the difference in temperature from the previous reference (i.e. $\Delta T$)}\label{fig:QualitativeResltTmpRecon}
\end{figure*}

The temperature difference between the ground truth and the highest undersampled images (2D varden 10\%) was 1.299±0.032, which is 1.3 ℃ more than the ground truth. But the models managed to reduce the difference to 0.618±0.016 and 0.643±0.022, using Fourier-PDNet and Fourier-PDUNet models, respectively, which are only around half a ℃ more than the ground truth (see Table \ref{tab:Quantativerslt}). So, the models give 60\% better accuracy in reconstructing the temperature maps compared to undersampled MRIs. This means that the model can speed up MR acquisition by a factor of 10 with only half ℃  of temperature difference. Examples of the reconstructed temperature map are shown in Fig. \ref{fig:QualitativeResltTmpRecon}.

Moreover, the temperature difference between the ground truth and the most undersampled images (2D Varden 10\%) was 0.488±0.161, which is 0.49°C higher than the ground truth. However, the Fourier-PDNet and Fourier-PDUNet models managed to reduce this difference to 0.063±0.009 and 0.11±0.026 , respectively, which is only around 0.10 °C Celsius above the ground truth. These scores are presented in Table \ref{tab:ROIQuantativerslt}.

\section{Discussion}
\noindent The assessment of the proposed framework for reconstructing MR images from undersampled data has shown that it can be applied effectively not only to MRI but also to MR thermometry images, as the model was also capable of reconstructing the temperatures. To the authors' best knowledge, this manuscript is the first one to deal with undersampled MR thermometry, and in extension, MR-guided hyperhermia, using deep learning-based methods; while this is also the first research discussing the need and possibility of accelerating MR-guided hypetrhermia. 

From the results, it can be observed that the framework seems to be robust against various undersampling patterns. For example, the SSIM values of 1D varden 25\% are 63\% and 31\%, for the magnitude and phase images where the Fourier-PDNet and Fourier-PDUNet models managed to reconstruct those data with average SSIM values of 91\% and 90\% for the magnitude images, while achieving 44\% and 40\% for the phase images, respectively.
SSIM values of the 2D varden 25\% are 43\% and 29\% for the magnitude and phase images. The Fourier-PDNet and Fourier-PDUNet models managed to reconstruct those data with average SSIM values of 94\% and 93\% for the magnitude images while achieving 47\% and 46\% for the phase images.
2D varden 10\% is 39\% and 28\% for the magnitude and the phase images. The Fourier-PDNet and Fourier-PDUNet models managed to reconstruct those data with average SSIM values of  87\% and 86\% for the magnitude images while achieving 43\% and 41\% for the phase images, respectively, and the result has been displayed in violin plot\ref{fig:NewQualitativeReslt1} as well as in table \ref{tab:ResultTable}

The results show that both the Fourier Primal-Dual network (PD Net) and Fourier Primal-Dual UNet (PDUNet) were able to alleviate the undersampling problem and show that the deep learning model has the potential to improve the novel hyperthermia treatment. 
From the quantitative result Table \ref{tab:ResultTable}, it has been clear that Fourier PD net outperformed Fourier PDUNet in SSIM.
The same phenomenon has been observed in UIQI, but a slightly different phenomenon has been reported for NRMSE. For the 1D varden 25\% and the 2D varden 25\%, the value of NRMSE for PDUNet outperformed the output PDNet. 

Also, from the result, it has been clear that both of the models performed way better for magnitude images than the phase images, which could be the reason for the present  0.5℃ temperature difference in the reconstructed temperature from the ground truth. Improvement of the models for phase images can also decrease the difference in temperature. It is worth mentioning that temperature differences were computed considering the fully-sampled data as the ground truth, and the accuracy of the temperature maps generated from fully-sampled data was not explored, which might have been affected by different factors, such as the $B_0$ drift, as that was outside the scope of this research.

Furthermore, it is important to consider that the proposed method currently functions with already coil-combined data, and the artificial undersampling technique employed here simulates only a single channel. Thus, to employ this in a clinical setting, zero-filled coil-combined data obtained directly from the scanner can be provided as input into this method for reconstruction. This may also facilitate further acceleration through the use of parallel imaging techniques.

Finally, this research demonstrates the possibility of accelerating MR thermometry during MR-guided hyperthermia. By using undersampling patterns such as 2D varden 10\%, the acquisition can be ten times faster, and the methods presented here reduce the compromise in terms of temperature accuracy considerably. The static MRIs used in this research took between 33 and 93 seconds (see Table \ref{tab:Static MR acquisition information}) to acquire the whole volume, while the dynamic acquisitions took 118 to 225 seconds (see Table \ref{tab:Dynamic MR acquisition information}). Reducing the scan duration with acceleration factors of 4 or 10, as presented here, can considerably reduce the probability of patient movements (voluntary or involuntary). Theoretically, the static scan time might be reduced to between 3 and 9 seconds, and the dynamic scan time to between 12 and 23 seconds with an acceleration factor of 10. Faster acquisition would not only reduce the chances of motion artefacts due to patient movements during the scan, but it would also significantly improve the temporal resolution of the imaging and enhance temperature tracking across time points by capturing subtle changes in temperature over time.

\section{Conclusion and future works}
\noindent This paper introduced deep learning-based reconstruction of undersampled MR thermometry acquired during hyperthermia using Fourier-PDNet and Fourier-PDUNet models. After all the different experiments with different types of undersampling methods of different percentages, the results show that the methods were able to alleviate the undersampling problem and managed to get SSIM Score of 0.886±0.004 for magnitude images and 0.429±0.01 for phase images for highest underampling pattern 2D varden 10\% - which means that the MR acquisition is now ten times faster and it also manages to bring the temperature difference close to the ground-truth which is 0.618±0.016 in the full volume and 0.063±0.009 in the tumour region. Still, half a ℃ temperature difference (in the full volume) can be seen in the deep learning results. This can be attributed to the performance difference of the models between the magnitude and phase images. 

Future work will focus on improving the networks' performance on the phase images, which should also reduce the temperature difference. Furthermore, combining the Fourier-PDNet and Fourier-PDUNet models with dynamic MRI-centric pipelines \citep{sarasaen2021fine,chatterjee2022ddos} could allow these models to better exploit the spatio-temporal nature of MR thermometry data, improving the overall reconstruction quality. This work utilised already coil-combined single-channel input and functions as a post-hoc technique. In future research, this method might be extended to an end-to-end framework by working directly with coil images.

Another future direction for research is to focus on the latent space. Exploring the latent space can be useful for improving the image reconstruction quality in undersampled image reconstruction tasks where the input images contain artefacts; the latent space represents, in theory, a low-dimensional representation of the input images without the artefacts. The input images with artefacts can be considered as augmented versions of the input images. Different types of variational auto-encoder\citep{makhzani2015adversarial} methods can be used, such as Factorised Variational Auto-encoder (FactorVAE) \citep{kim2018disentangling}, Vector Quantised Variational Auto-encoder (VQ-VAE) \citep{van2017neural}, Masked autoencoders (MAE)\citep{He_2022_CVPR} etc.
Use of post hoc explainability methods like Saliency \citep{simonyan2013deep}, Occlusion \citep{zeiler2014visualizing}, Guided Backpropagation \citep{mahendran2016salient} etc. can give a better understanding of what went wrong with phase images, which can help the authors to improve the network accordingly. 

\section*{Data availability}
The datasets used and/or analysed during the current study available from the senior authors (U.O.G.: udo.gaipl@uk-erlangen.de and B.F.: benjamin.frey@uk-erlangen.de) on reasonable request, following the data privacy policy of Universitätsklinikum Erlangen.

\section *{Acknowledgement}
This research has received support from the European Union’s Horizon 2020 research and
innovation programme under the Marie Skłodowska-Curie (MSCA-ITN) grant “Hyperboost” project,
no. 955625. The present work was performed by Rupali Khatun in (partial) fulfilment of the requirements for obtaining the degree "Dr. rer. biol. hum." at the Friedrich-Alexnder-Universität Erlangen-Nürnberg (FAU).





\bibliography{mybibfile}

\begin{thebibliography}{54}
\expandafter\ifx\csname natexlab\endcsname\relax\def\natexlab#1{#1}\fi
\providecommand{\url}[1]{\texttt{#1}}
\providecommand{\href}[2]{#2}
\providecommand{\path}[1]{#1}
\providecommand{\DOIprefix}{doi:}
\providecommand{\ArXivprefix}{arXiv:}
\providecommand{\URLprefix}{URL: }
\providecommand{\Pubmedprefix}{pmid:}
\providecommand{\doi}[1]{\href{http://dx.doi.org/#1}{\path{#1}}}
\providecommand{\Pubmed}[1]{\href{pmid:#1}{\path{#1}}}
\providecommand{\bibinfo}[2]{#2}
\ifx\xfnm\relax \def\xfnm[#1]{\unskip,\space#1}\fi
\bibitem[{Adler and {\"O}ktem(2018)}]{adler2018learned}
\bibinfo{author}{Adler, J.}, \bibinfo{author}{{\"O}ktem, O.},
  \bibinfo{year}{2018}.
\newblock \bibinfo{title}{Learned primal-dual reconstruction}.
\newblock \bibinfo{journal}{IEEE transactions on medical imaging}
  \bibinfo{volume}{37}, \bibinfo{pages}{1322--1332}.
\bibitem[{Cernicanu et~al.(2008)Cernicanu, Lepetit-Coiffe, Roland, Becker and
  Terraz}]{cernicanu2008validation}
\bibinfo{author}{Cernicanu, A.}, \bibinfo{author}{Lepetit-Coiffe, M.},
  \bibinfo{author}{Roland, J.}, \bibinfo{author}{Becker, C.D.},
  \bibinfo{author}{Terraz, S.}, \bibinfo{year}{2008}.
\newblock \bibinfo{title}{Validation of fast mr thermometry at 1.5 t with
  gradient-echo echo planar imaging sequences: phantom and clinical feasibility
  studies}.
\newblock \bibinfo{journal}{NMR in Biomedicine: An International Journal
  Devoted to the Development and Application of Magnetic Resonance In vivo}
  \bibinfo{volume}{21}, \bibinfo{pages}{849--858}.
\bibitem[{Chatterjee(2020)}]{chatterjee2020soumickmj}
\bibinfo{author}{Chatterjee, S.}, \bibinfo{year}{2020}.
\newblock \bibinfo{title}{soumickmj/mrunder: Initial release (version v0. 1)}.
\newblock \bibinfo{journal}{DOI: http://doi. org/10.5281/zenodo}
  \bibinfo{volume}{3901455}.
\bibitem[{Chatterjee et~al.(2022a)Chatterjee, Breitkopf, Sarasaen, Yassin,
  Rose, N{\"u}rnberger and Speck}]{chatterjee2022reconresnet}
\bibinfo{author}{Chatterjee, S.}, \bibinfo{author}{Breitkopf, M.},
  \bibinfo{author}{Sarasaen, C.}, \bibinfo{author}{Yassin, H.},
  \bibinfo{author}{Rose, G.}, \bibinfo{author}{N{\"u}rnberger, A.},
  \bibinfo{author}{Speck, O.}, \bibinfo{year}{2022}a.
\newblock \bibinfo{title}{Reconresnet: Regularised residual learning for mr
  image reconstruction of undersampled cartesian and radial data}.
\newblock \bibinfo{journal}{Computers in Biology and Medicine}
  \bibinfo{volume}{143}, \bibinfo{pages}{105321}.
\bibitem[{Chatterjee et~al.(2024)Chatterjee, Sarasaen, Rose, Nürnberger and
  Speck}]{chatterjee2022ddos}
\bibinfo{author}{Chatterjee, S.}, \bibinfo{author}{Sarasaen, C.},
  \bibinfo{author}{Rose, G.}, \bibinfo{author}{Nürnberger, A.},
  \bibinfo{author}{Speck, O.}, \bibinfo{year}{2024}.
\newblock \bibinfo{title}{Ddos-unet: Incorporating temporal information using
  dynamic dual-channel unet for enhancing super-resolution of dynamic mri}.
\newblock \bibinfo{journal}{IEEE Access}
  \DOIprefix\doi{10.1109/ACCESS.2024.3427674}.
\bibitem[{Chatterjee et~al.(2021)Chatterjee, Sarasaen, Sciarra, Breitkopf,
  Oeltze-Jafra, N{\"u}rnberger and Speck}]{chatterjee2021going}
\bibinfo{author}{Chatterjee, S.}, \bibinfo{author}{Sarasaen, C.},
  \bibinfo{author}{Sciarra, A.}, \bibinfo{author}{Breitkopf, M.},
  \bibinfo{author}{Oeltze-Jafra, S.}, \bibinfo{author}{N{\"u}rnberger, A.},
  \bibinfo{author}{Speck, O.}, \bibinfo{year}{2021}.
\newblock \bibinfo{title}{Going beyond the image space: undersampled mri
  reconstruction directly in the k-space using a complex valued residual neural
  network}, in: \bibinfo{booktitle}{2021 ISMRM \& SMRT Annual Meeting \&
  Exhibition}, p. \bibinfo{pages}{1757}.
\bibitem[{Chatterjee et~al.(2022b)Chatterjee, Tummala, Speck and
  N{\"u}rnberger}]{chatterjee2022complex}
\bibinfo{author}{Chatterjee, S.}, \bibinfo{author}{Tummala, P.},
  \bibinfo{author}{Speck, O.}, \bibinfo{author}{N{\"u}rnberger, A.},
  \bibinfo{year}{2022}b.
\newblock \bibinfo{title}{Complex network for complex problems: A comparative
  study of cnn and complex-valued cnn}, in: \bibinfo{booktitle}{2022 IEEE 5th
  International Conference on Image Processing Applications and Systems
  (IPAS)}, \bibinfo{organization}{IEEE}. pp. \bibinfo{pages}{1--5}.
\bibitem[{Cihoric et~al.(2015)Cihoric, Tsikkinis, van Rhoon, Crezee, Aebersold,
  Bodis, Beck, Nadobny, Budach, Wust et~al.}]{cihoric2015hyperthermia}
\bibinfo{author}{Cihoric, N.}, \bibinfo{author}{Tsikkinis, A.},
  \bibinfo{author}{van Rhoon, G.}, \bibinfo{author}{Crezee, H.},
  \bibinfo{author}{Aebersold, D.M.}, \bibinfo{author}{Bodis, S.},
  \bibinfo{author}{Beck, M.}, \bibinfo{author}{Nadobny, J.},
  \bibinfo{author}{Budach, V.}, \bibinfo{author}{Wust, P.}, et~al.,
  \bibinfo{year}{2015}.
\newblock \bibinfo{title}{Hyperthermia-related clinical trials on cancer
  treatment within the clinicaltrials. gov registry}.
\newblock \bibinfo{journal}{International journal of hyperthermia}
  \bibinfo{volume}{31}, \bibinfo{pages}{609--614}.
\bibitem[{Cline et~al.(1994)Cline, Hynynen, Hardy, Watkins, Schenck and
  Jolesz}]{cline1994mr}
\bibinfo{author}{Cline, H.E.}, \bibinfo{author}{Hynynen, K.},
  \bibinfo{author}{Hardy, C.J.}, \bibinfo{author}{Watkins, R.D.},
  \bibinfo{author}{Schenck, J.F.}, \bibinfo{author}{Jolesz, F.A.},
  \bibinfo{year}{1994}.
\newblock \bibinfo{title}{Mr temperature mapping of focused ultrasound
  surgery}.
\newblock \bibinfo{journal}{Magnetic resonance in medicine}
  \bibinfo{volume}{31}, \bibinfo{pages}{628--636}.
\bibitem[{Datta et~al.(2019)Datta, Stutz, Gomez and Bodis}]{datta2019efficacy}
\bibinfo{author}{Datta, N.R.}, \bibinfo{author}{Stutz, E.},
  \bibinfo{author}{Gomez, S.}, \bibinfo{author}{Bodis, S.},
  \bibinfo{year}{2019}.
\newblock \bibinfo{title}{Efficacy and safety evaluation of the various
  therapeutic options in locally advanced cervix cancer: a systematic review
  and network meta-analysis of randomized clinical trials}.
\newblock \bibinfo{journal}{International Journal of Radiation Oncology*
  Biology* Physics} \bibinfo{volume}{103}, \bibinfo{pages}{411--437}.
\bibitem[{Ernst et~al.(2023)Ernst, Chatterjee, Rose, Speck and
  N{\"u}rnberger}]{ernst2021sinogram}
\bibinfo{author}{Ernst, P.}, \bibinfo{author}{Chatterjee, S.},
  \bibinfo{author}{Rose, G.}, \bibinfo{author}{Speck, O.},
  \bibinfo{author}{N{\"u}rnberger, A.}, \bibinfo{year}{2023}.
\newblock \bibinfo{title}{Sinogram upsampling using primal--dual unet for
  undersampled ct and radial mri reconstruction}.
\newblock \bibinfo{journal}{Neural Networks} \bibinfo{volume}{166},
  \bibinfo{pages}{704--721}.
\bibitem[{Gellermann et~al.(2005)Gellermann, Wlodarczyk, Feussner, F{\"a}hling,
  Nadobny, Hildebrandt, Felix and Wust}]{gellermann2005methods}
\bibinfo{author}{Gellermann, J.}, \bibinfo{author}{Wlodarczyk, W.},
  \bibinfo{author}{Feussner, A.}, \bibinfo{author}{F{\"a}hling, H.},
  \bibinfo{author}{Nadobny, J.}, \bibinfo{author}{Hildebrandt, B.},
  \bibinfo{author}{Felix, R.}, \bibinfo{author}{Wust, P.},
  \bibinfo{year}{2005}.
\newblock \bibinfo{title}{Methods and potentials of magnetic resonance imaging
  for monitoring radiofrequency hyperthermia in a hybrid system}.
\newblock \bibinfo{journal}{International journal of hyperthermia}
  \bibinfo{volume}{21}, \bibinfo{pages}{497--513}.
\bibitem[{He et~al.(2022)He, Chen, Xie, Li, Doll\'ar and
  Girshick}]{He_2022_CVPR}
\bibinfo{author}{He, K.}, \bibinfo{author}{Chen, X.}, \bibinfo{author}{Xie,
  S.}, \bibinfo{author}{Li, Y.}, \bibinfo{author}{Doll\'ar, P.},
  \bibinfo{author}{Girshick, R.}, \bibinfo{year}{2022}.
\newblock \bibinfo{title}{Masked autoencoders are scalable vision learners},
  in: \bibinfo{booktitle}{Proceedings of the IEEE/CVF Conference on Computer
  Vision and Pattern Recognition (CVPR)}, pp. \bibinfo{pages}{16000--16009}.
\bibitem[{He et~al.(2016)He, Zhang, Ren and Sun}]{he2016deep}
\bibinfo{author}{He, K.}, \bibinfo{author}{Zhang, X.}, \bibinfo{author}{Ren,
  S.}, \bibinfo{author}{Sun, J.}, \bibinfo{year}{2016}.
\newblock \bibinfo{title}{Deep residual learning for image recognition}, in:
  \bibinfo{booktitle}{Proceedings of the IEEE conference on computer vision and
  pattern recognition}, pp. \bibinfo{pages}{770--778}.
\bibitem[{Hyun et~al.(2018)Hyun, Kim, Lee, Lee and Seo}]{hyun2018deep}
\bibinfo{author}{Hyun, C.M.}, \bibinfo{author}{Kim, H.P.},
  \bibinfo{author}{Lee, S.M.}, \bibinfo{author}{Lee, S.}, \bibinfo{author}{Seo,
  J.K.}, \bibinfo{year}{2018}.
\newblock \bibinfo{title}{Deep learning for undersampled mri reconstruction}.
\newblock \bibinfo{journal}{Physics in Medicine \& Biology}
  \bibinfo{volume}{63}, \bibinfo{pages}{135007}.
\bibitem[{Ishihara et~al.(1995)Ishihara, Calderon, Watanabe, Okamoto, Suzuki,
  Kuroda and Suzuki}]{ishihara1995precise}
\bibinfo{author}{Ishihara, Y.}, \bibinfo{author}{Calderon, A.},
  \bibinfo{author}{Watanabe, H.}, \bibinfo{author}{Okamoto, K.},
  \bibinfo{author}{Suzuki, Y.}, \bibinfo{author}{Kuroda, K.},
  \bibinfo{author}{Suzuki, Y.}, \bibinfo{year}{1995}.
\newblock \bibinfo{title}{A precise and fast temperature mapping using water
  proton chemical shift}.
\newblock \bibinfo{journal}{Magnetic resonance in medicine}
  \bibinfo{volume}{34}, \bibinfo{pages}{814--823}.
\bibitem[{Jifara et~al.(2019)Jifara, Jiang, Rho, Cheng and
  Liu}]{jifara2019medical}
\bibinfo{author}{Jifara, W.}, \bibinfo{author}{Jiang, F.},
  \bibinfo{author}{Rho, S.}, \bibinfo{author}{Cheng, M.}, \bibinfo{author}{Liu,
  S.}, \bibinfo{year}{2019}.
\newblock \bibinfo{title}{Medical image denoising using convolutional neural
  network: a residual learning approach}.
\newblock \bibinfo{journal}{The Journal of Supercomputing}
  \bibinfo{volume}{75}, \bibinfo{pages}{704--718}.
\bibitem[{Kim and Mnih(2018)}]{kim2018disentangling}
\bibinfo{author}{Kim, H.}, \bibinfo{author}{Mnih, A.}, \bibinfo{year}{2018}.
\newblock \bibinfo{title}{Disentangling by factorising}, in:
  \bibinfo{booktitle}{International Conference on Machine Learning},
  \bibinfo{organization}{PMLR}. pp. \bibinfo{pages}{2649--2658}.
\bibitem[{Kim and Hahn(1979)}]{kim1979clinical}
\bibinfo{author}{Kim, J.}, \bibinfo{author}{Hahn, E.}, \bibinfo{year}{1979}.
\newblock \bibinfo{title}{Clinical and biological studies of localized
  hyperthermia}.
\newblock \bibinfo{journal}{Cancer Research} \bibinfo{volume}{39},
  \bibinfo{pages}{2258--2261}.
\bibitem[{Kok et~al.(2020)Kok, Beck, L{\"o}ke, Helderman, van Tienhoven,
  Ghadjar, Wust and Crezee}]{kok2020locoregional}
\bibinfo{author}{Kok, H.P.}, \bibinfo{author}{Beck, M.},
  \bibinfo{author}{L{\"o}ke, D.R.}, \bibinfo{author}{Helderman, R.F.},
  \bibinfo{author}{van Tienhoven, G.}, \bibinfo{author}{Ghadjar, P.},
  \bibinfo{author}{Wust, P.}, \bibinfo{author}{Crezee, H.},
  \bibinfo{year}{2020}.
\newblock \bibinfo{title}{Locoregional peritoneal hyperthermia to enhance the
  effectiveness of chemotherapy in patients with peritoneal carcinomatosis: A
  simulation study comparing different locoregional heating systems}.
\newblock \bibinfo{journal}{International Journal of Hyperthermia}
  \bibinfo{volume}{37}, \bibinfo{pages}{76--88}.
\bibitem[{Kuroda(2005)}]{kuroda2005non}
\bibinfo{author}{Kuroda, K.}, \bibinfo{year}{2005}.
\newblock \bibinfo{title}{Non-invasive mr thermography using the water proton
  chemical shift}.
\newblock \bibinfo{journal}{International journal of hyperthermia}
  \bibinfo{volume}{21}, \bibinfo{pages}{547--560}.
\bibitem[{L{\"u}demann et~al.(2010)L{\"u}demann, Wlodarczyk, Nadobny,
  Weihrauch, Gellermann and Wust}]{ludemann2010non}
\bibinfo{author}{L{\"u}demann, L.}, \bibinfo{author}{Wlodarczyk, W.},
  \bibinfo{author}{Nadobny, J.}, \bibinfo{author}{Weihrauch, M.},
  \bibinfo{author}{Gellermann, J.}, \bibinfo{author}{Wust, P.},
  \bibinfo{year}{2010}.
\newblock \bibinfo{title}{Non-invasive magnetic resonance thermography during
  regional hyperthermia}.
\newblock \bibinfo{journal}{International Journal of Hyperthermia}
  \bibinfo{volume}{26}, \bibinfo{pages}{273--282}.
\bibitem[{Lustig et~al.(2007)Lustig, Donoho and Pauly}]{lustig2007sparse}
\bibinfo{author}{Lustig, M.}, \bibinfo{author}{Donoho, D.},
  \bibinfo{author}{Pauly, J.M.}, \bibinfo{year}{2007}.
\newblock \bibinfo{title}{Sparse mri: The application of compressed sensing for
  rapid mr imaging}.
\newblock \bibinfo{journal}{Magnetic Resonance in Medicine: An Official Journal
  of the International Society for Magnetic Resonance in Medicine}
  \bibinfo{volume}{58}, \bibinfo{pages}{1182--1195}.
\bibitem[{Lyu et~al.(2021)Lyu, Shan, Xie, Kwan, Otaki, Kuronuma, Li and
  Wang}]{lyu2021cine}
\bibinfo{author}{Lyu, Q.}, \bibinfo{author}{Shan, H.}, \bibinfo{author}{Xie,
  Y.}, \bibinfo{author}{Kwan, A.C.}, \bibinfo{author}{Otaki, Y.},
  \bibinfo{author}{Kuronuma, K.}, \bibinfo{author}{Li, D.},
  \bibinfo{author}{Wang, G.}, \bibinfo{year}{2021}.
\newblock \bibinfo{title}{Cine cardiac mri motion artifact reduction using a
  recurrent neural network}.
\newblock \bibinfo{journal}{IEEE Transactions on Medical Imaging}
  \bibinfo{volume}{40}, \bibinfo{pages}{2170--2181}.
\bibitem[{Madore et~al.(2011)Madore, Panych, Mei, Yuan and
  Chu}]{madore2011multipathway}
\bibinfo{author}{Madore, B.}, \bibinfo{author}{Panych, L.P.},
  \bibinfo{author}{Mei, C.S.}, \bibinfo{author}{Yuan, J.},
  \bibinfo{author}{Chu, R.}, \bibinfo{year}{2011}.
\newblock \bibinfo{title}{Multipathway sequences for mr thermometry}.
\newblock \bibinfo{journal}{Magnetic resonance in medicine}
  \bibinfo{volume}{66}, \bibinfo{pages}{658--668}.
\bibitem[{Mahendran and Vedaldi(2016)}]{mahendran2016salient}
\bibinfo{author}{Mahendran, A.}, \bibinfo{author}{Vedaldi, A.},
  \bibinfo{year}{2016}.
\newblock \bibinfo{title}{Salient deconvolutional networks}, in:
  \bibinfo{booktitle}{European Conference on Computer Vision},
  \bibinfo{organization}{Springer}. pp. \bibinfo{pages}{120--135}.
\bibitem[{Makhzani et~al.(2015)Makhzani, Shlens, Jaitly, Goodfellow and
  Frey}]{makhzani2015adversarial}
\bibinfo{author}{Makhzani, A.}, \bibinfo{author}{Shlens, J.},
  \bibinfo{author}{Jaitly, N.}, \bibinfo{author}{Goodfellow, I.},
  \bibinfo{author}{Frey, B.}, \bibinfo{year}{2015}.
\newblock \bibinfo{title}{Adversarial autoencoders}.
\newblock \bibinfo{journal}{arXiv preprint arXiv:1511.05644} .
\bibitem[{McDannold(2005)}]{mcdannold2005quantitative}
\bibinfo{author}{McDannold, N.}, \bibinfo{year}{2005}.
\newblock \bibinfo{title}{Quantitative mri-based temperature mapping based on
  the proton resonant frequency shift: review of validation studies}.
\newblock \bibinfo{journal}{International journal of hyperthermia}
  \bibinfo{volume}{21}, \bibinfo{pages}{533--546}.
\bibitem[{Mou et~al.(2017)Mou, Ghamisi and Zhu}]{mou2017unsupervised}
\bibinfo{author}{Mou, L.}, \bibinfo{author}{Ghamisi, P.}, \bibinfo{author}{Zhu,
  X.X.}, \bibinfo{year}{2017}.
\newblock \bibinfo{title}{Unsupervised spectral--spatial feature learning via
  deep residual conv--deconv network for hyperspectral image classification}.
\newblock \bibinfo{journal}{IEEE Transactions on Geoscience and Remote Sensing}
  \bibinfo{volume}{56}, \bibinfo{pages}{391--406}.
\bibitem[{Nyquist(1928)}]{nyquist1928certain}
\bibinfo{author}{Nyquist, H.}, \bibinfo{year}{1928}.
\newblock \bibinfo{title}{Certain topics in telegraph transmission theory}.
\newblock \bibinfo{journal}{Transactions of the American Institute of
  Electrical Engineers} \bibinfo{volume}{47}, \bibinfo{pages}{617--644}.
\bibitem[{Od{\'e}en and Parker(2019)}]{odeen2019magnetic}
\bibinfo{author}{Od{\'e}en, H.}, \bibinfo{author}{Parker, D.L.},
  \bibinfo{year}{2019}.
\newblock \bibinfo{title}{Magnetic resonance thermometry and its biological
  applications--physical principles and practical considerations}.
\newblock \bibinfo{journal}{Progress in nuclear magnetic resonance
  spectroscopy} \bibinfo{volume}{110}, \bibinfo{pages}{34--61}.
\bibitem[{Pakhomov et~al.(2019)Pakhomov, Premachandran, Allan, Azizian and
  Navab}]{pakhomov2019deep}
\bibinfo{author}{Pakhomov, D.}, \bibinfo{author}{Premachandran, V.},
  \bibinfo{author}{Allan, M.}, \bibinfo{author}{Azizian, M.},
  \bibinfo{author}{Navab, N.}, \bibinfo{year}{2019}.
\newblock \bibinfo{title}{Deep residual learning for instrument segmentation in
  robotic surgery}, in: \bibinfo{booktitle}{International Workshop on Machine
  Learning in Medical Imaging}, \bibinfo{organization}{Springer}. pp.
  \bibinfo{pages}{566--573}.
\bibitem[{Parker et~al.(1983)Parker, Smith, Sheldon, Crooks and
  Fussell}]{parker1983temperature}
\bibinfo{author}{Parker, D.L.}, \bibinfo{author}{Smith, V.},
  \bibinfo{author}{Sheldon, P.}, \bibinfo{author}{Crooks, L.E.},
  \bibinfo{author}{Fussell, L.}, \bibinfo{year}{1983}.
\newblock \bibinfo{title}{Temperature distribution measurements in
  two-dimensional nmr imaging}.
\newblock \bibinfo{journal}{Medical physics} \bibinfo{volume}{10},
  \bibinfo{pages}{321--325}.
\bibitem[{Paszke et~al.(2019)Paszke, Gross, Massa, Lerer, Bradbury, Chanan,
  Killeen, Lin, Gimelshein, Antiga et~al.}]{paszke2019pytorch}
\bibinfo{author}{Paszke, A.}, \bibinfo{author}{Gross, S.},
  \bibinfo{author}{Massa, F.}, \bibinfo{author}{Lerer, A.},
  \bibinfo{author}{Bradbury, J.}, \bibinfo{author}{Chanan, G.},
  \bibinfo{author}{Killeen, T.}, \bibinfo{author}{Lin, Z.},
  \bibinfo{author}{Gimelshein, N.}, \bibinfo{author}{Antiga, L.}, et~al.,
  \bibinfo{year}{2019}.
\newblock \bibinfo{title}{Pytorch: An imperative style, high-performance deep
  learning library}.
\newblock \bibinfo{journal}{Advances in neural information processing systems}
  \bibinfo{volume}{32}.
\bibitem[{Peters(2000)}]{peters2000magnetic}
\bibinfo{author}{Peters, R.D.}, \bibinfo{year}{2000}.
\newblock \bibinfo{title}{Magnetic resonance thermometry for image-guided
  thermal therapy}.
\newblock \bibinfo{journal}{Toronto: University of Toronto} .
\bibitem[{Qin et~al.(2018)Qin, Schlemper, Caballero, Price, Hajnal and
  Rueckert}]{qin2018convolutional}
\bibinfo{author}{Qin, C.}, \bibinfo{author}{Schlemper, J.},
  \bibinfo{author}{Caballero, J.}, \bibinfo{author}{Price, A.N.},
  \bibinfo{author}{Hajnal, J.V.}, \bibinfo{author}{Rueckert, D.},
  \bibinfo{year}{2018}.
\newblock \bibinfo{title}{Convolutional recurrent neural networks for dynamic
  mr image reconstruction}.
\newblock \bibinfo{journal}{IEEE transactions on medical imaging}
  \bibinfo{volume}{38}, \bibinfo{pages}{280--290}.
\bibitem[{Quesson et~al.(2000)Quesson, de~Zwart and
  Moonen}]{quesson2000magnetic}
\bibinfo{author}{Quesson, B.}, \bibinfo{author}{de~Zwart, J.A.},
  \bibinfo{author}{Moonen, C.T.}, \bibinfo{year}{2000}.
\newblock \bibinfo{title}{Magnetic resonance temperature imaging for guidance
  of thermotherapy}.
\newblock \bibinfo{journal}{Journal of Magnetic Resonance Imaging: An Official
  Journal of the International Society for Magnetic Resonance in Medicine}
  \bibinfo{volume}{12}, \bibinfo{pages}{525--533}.
\bibitem[{Renieblas et~al.(2017)Renieblas, Nogu{\'e}s, Gonz{\'a}lez, Le{\'o}n
  and Del~Castillo}]{renieblas2017structural}
\bibinfo{author}{Renieblas, G.P.}, \bibinfo{author}{Nogu{\'e}s, A.T.},
  \bibinfo{author}{Gonz{\'a}lez, A.M.}, \bibinfo{author}{Le{\'o}n, N.G.},
  \bibinfo{author}{Del~Castillo, E.G.}, \bibinfo{year}{2017}.
\newblock \bibinfo{title}{Structural similarity index family for image quality
  assessment in radiological images}.
\newblock \bibinfo{journal}{Journal of medical imaging} \bibinfo{volume}{4},
  \bibinfo{pages}{035501}.
\bibitem[{Rieke and Butts~Pauly(2008)}]{rieke2008mr}
\bibinfo{author}{Rieke, V.}, \bibinfo{author}{Butts~Pauly, K.},
  \bibinfo{year}{2008}.
\newblock \bibinfo{title}{Mr thermometry}.
\newblock \bibinfo{journal}{Journal of Magnetic Resonance Imaging: An Official
  Journal of the International Society for Magnetic Resonance in Medicine}
  \bibinfo{volume}{27}, \bibinfo{pages}{376--390}.
\bibitem[{Rieke and Pauly(2008)}]{rieke2008journal}
\bibinfo{author}{Rieke, V.}, \bibinfo{author}{Pauly, K.B.},
  \bibinfo{year}{2008}.
\newblock \bibinfo{title}{Journal of magnetic resonance imaging: Jmri}.
\newblock \bibinfo{journal}{J. Magn. Reson. Imag.} \bibinfo{volume}{27},
  \bibinfo{pages}{376--390}.
\bibitem[{Rossmann and Haemmerich(2014)}]{rossmann2014review}
\bibinfo{author}{Rossmann, C.}, \bibinfo{author}{Haemmerich, D.},
  \bibinfo{year}{2014}.
\newblock \bibinfo{title}{Review of temperature dependence of thermal
  properties, dielectric properties, and perfusion of biological tissues at
  hyperthermic and ablation temperatures}.
\newblock \bibinfo{journal}{Critical Reviews™ in Biomedical Engineering}
  \bibinfo{volume}{42}.
\bibitem[{Sarasaen et~al.(2021)Sarasaen, Chatterjee, Breitkopf, Rose,
  N{\"u}rnberger and Speck}]{sarasaen2021fine}
\bibinfo{author}{Sarasaen, C.}, \bibinfo{author}{Chatterjee, S.},
  \bibinfo{author}{Breitkopf, M.}, \bibinfo{author}{Rose, G.},
  \bibinfo{author}{N{\"u}rnberger, A.}, \bibinfo{author}{Speck, O.},
  \bibinfo{year}{2021}.
\newblock \bibinfo{title}{Fine-tuning deep learning model parameters for
  improved super-resolution of dynamic mri with prior-knowledge}.
\newblock \bibinfo{journal}{Artificial Intelligence in Medicine}
  \bibinfo{volume}{121}, \bibinfo{pages}{102196}.
\bibitem[{Shannon(1949)}]{shannon1949communication}
\bibinfo{author}{Shannon, C.E.}, \bibinfo{year}{1949}.
\newblock \bibinfo{title}{Communication in the presence of noise}.
\newblock \bibinfo{journal}{Proceedings of the IRE} \bibinfo{volume}{37},
  \bibinfo{pages}{10--21}.
\bibitem[{Simonyan et~al.(2013)Simonyan, Vedaldi and
  Zisserman}]{simonyan2013deep}
\bibinfo{author}{Simonyan, K.}, \bibinfo{author}{Vedaldi, A.},
  \bibinfo{author}{Zisserman, A.}, \bibinfo{year}{2013}.
\newblock \bibinfo{title}{Deep inside convolutional networks: Visualising image
  classification models and saliency maps}.
\newblock \bibinfo{journal}{arXiv preprint arXiv:1312.6034} .
\bibitem[{Thomsen(1991)}]{thomsen1991pathologic}
\bibinfo{author}{Thomsen, S.}, \bibinfo{year}{1991}.
\newblock \bibinfo{title}{Pathologic analysis of photothermal and
  photomechanical effects of laser--tissue interactions}.
\newblock \bibinfo{journal}{Photochemistry and photobiology}
  \bibinfo{volume}{53}, \bibinfo{pages}{825--835}.
\bibitem[{Van Den~Oord et~al.(2017)Van Den~Oord, Vinyals
  et~al.}]{van2017neural}
\bibinfo{author}{Van Den~Oord, A.}, \bibinfo{author}{Vinyals, O.}, et~al.,
  \bibinfo{year}{2017}.
\newblock \bibinfo{title}{Neural discrete representation learning}.
\newblock \bibinfo{journal}{Advances in neural information processing systems}
  \bibinfo{volume}{30}.
\bibitem[{Wang et~al.(2016)Wang, Su, Ying, Peng, Zhu, Liang, Feng and
  Liang}]{wang2016accelerating}
\bibinfo{author}{Wang, S.}, \bibinfo{author}{Su, Z.}, \bibinfo{author}{Ying,
  L.}, \bibinfo{author}{Peng, X.}, \bibinfo{author}{Zhu, S.},
  \bibinfo{author}{Liang, F.}, \bibinfo{author}{Feng, D.},
  \bibinfo{author}{Liang, D.}, \bibinfo{year}{2016}.
\newblock \bibinfo{title}{Accelerating magnetic resonance imaging via deep
  learning}, in: \bibinfo{booktitle}{2016 IEEE 13th international symposium on
  biomedical imaging (ISBI)}, \bibinfo{organization}{IEEE}. pp.
  \bibinfo{pages}{514--517}.
\bibitem[{Wang and Bovik(2002)}]{wang2002universal}
\bibinfo{author}{Wang, Z.}, \bibinfo{author}{Bovik, A.C.},
  \bibinfo{year}{2002}.
\newblock \bibinfo{title}{A universal image quality index}.
\newblock \bibinfo{journal}{IEEE signal processing letters}
  \bibinfo{volume}{9}, \bibinfo{pages}{81--84}.
\bibitem[{Wlodarczyk et~al.(1999)Wlodarczyk, Hentschel, Wust, Noeske, Hosten,
  Rinneberg and Felix}]{wlodarczyk1999comparison}
\bibinfo{author}{Wlodarczyk, W.}, \bibinfo{author}{Hentschel, M.},
  \bibinfo{author}{Wust, P.}, \bibinfo{author}{Noeske, R.},
  \bibinfo{author}{Hosten, N.}, \bibinfo{author}{Rinneberg, H.},
  \bibinfo{author}{Felix, R.}, \bibinfo{year}{1999}.
\newblock \bibinfo{title}{Comparison of four magnetic resonance methods for
  mapping small temperature changes}.
\newblock \bibinfo{journal}{Physics in Medicine \& Biology}
  \bibinfo{volume}{44}, \bibinfo{pages}{607}.
\bibitem[{Wust et~al.(2006)Wust, Cho, Hildebrandt and
  Gellermann}]{wust2006thermal}
\bibinfo{author}{Wust, P.}, \bibinfo{author}{Cho, C.H.},
  \bibinfo{author}{Hildebrandt, B.}, \bibinfo{author}{Gellermann, J.},
  \bibinfo{year}{2006}.
\newblock \bibinfo{title}{Thermal monitoring: invasive, minimal-invasive and
  non-invasive approaches}.
\newblock \bibinfo{journal}{International Journal of Hyperthermia}
  \bibinfo{volume}{22}, \bibinfo{pages}{255--262}.
\bibitem[{Van~der Zee(2002)}]{van2002heating}
\bibinfo{author}{Van~der Zee, J.}, \bibinfo{year}{2002}.
\newblock \bibinfo{title}{Heating the patient: a promising approach?}
\newblock \bibinfo{journal}{Annals of oncology} \bibinfo{volume}{13},
  \bibinfo{pages}{1173--1184}.
\bibitem[{Zeiler and Fergus(2014)}]{zeiler2014visualizing}
\bibinfo{author}{Zeiler, M.D.}, \bibinfo{author}{Fergus, R.},
  \bibinfo{year}{2014}.
\newblock \bibinfo{title}{Visualizing and understanding convolutional
  networks}, in: \bibinfo{booktitle}{European conference on computer vision},
  \bibinfo{organization}{Springer}. pp. \bibinfo{pages}{818--833}.
\bibitem[{Zhang et~al.(2019)Zhang, Xie, Xia and Shen}]{zhang2019attention}
\bibinfo{author}{Zhang, J.}, \bibinfo{author}{Xie, Y.}, \bibinfo{author}{Xia,
  Y.}, \bibinfo{author}{Shen, C.}, \bibinfo{year}{2019}.
\newblock \bibinfo{title}{Attention residual learning for skin lesion
  classification}.
\newblock \bibinfo{journal}{IEEE transactions on medical imaging}
  \bibinfo{volume}{38}, \bibinfo{pages}{2092--2103}.
\bibitem[{Zhu et~al.(2017)Zhu, Park, Isola and Efros}]{zhu2017unpaired}
\bibinfo{author}{Zhu, J.Y.}, \bibinfo{author}{Park, T.},
  \bibinfo{author}{Isola, P.}, \bibinfo{author}{Efros, A.A.},
  \bibinfo{year}{2017}.
\newblock \bibinfo{title}{Unpaired image-to-image translation using
  cycle-consistent adversarial networks}, in: \bibinfo{booktitle}{Proceedings
  of the IEEE international conference on computer vision}, pp.
  \bibinfo{pages}{2223--2232}.

\end{thebibliography}

\begin{table*}[hbtp!]
\begin{center}
\caption{Result of Cartesian undersampling patterns, while being trained separately}\label{tab:ResultTable}
\resizebox{\textwidth}{!}{%
\begin{tabular}{@{}cccccccccccc@{}}
\toprule
\multicolumn{3}{c}{\multirow{2}{*}{}}                                                   & \multicolumn{3}{c}{\textbf{2D Varden 10\%}}                                                   & \multicolumn{3}{c}{\textbf{2D Varden 25\%}}                                                   & \multicolumn{3}{c}{\textbf{1D Varden 25\%}}                                                   \\ \cmidrule(l){4-12} 
\multicolumn{3}{c}{}                                                                    & SSIM                          & NRMSE                         & UIQI                          & SSIM                          & NRMSE                         & UIQI                          & SSIM                          & NRMSE                         & UIQI                          \\ \midrule
\multirow{2}{*}{\textbf{Input}}  & \multirow{2}{*}{Undersample}    & \textit{Magnitude} & 0.385$\pm$0.009               & 0.514$\pm$0.022               & 0.536 $\pm$0.003              & 0.43±0.008                    & 0.478±0.015                   & 0.563±0.002                   & 0.635±0.005                   & 0.358±0.012                   & 0.705±0.003                   \\
                                 &                                 & \textit{Phase}     & 0.28$\pm$0.011                & 1.147$\pm$0.006               & 0.356 $\pm$0.013              & 0.29±0.011                    & 1.128±0.005                   & 0.363±0.012                   & 0.312±0.009                   & 1.072±0.005                   & 0.382±0.011                   \\
\multirow{4}{*}{\textbf{Output}} & \multirow{2}{*}{Fourier-PDNet}  & \textit{Magnitude} & \textit{\textbf{0.886±0.004}} & \textit{\textbf{0.121±0.002}} & \textit{\textbf{0.807±0.006}} & \textit{\textbf{0.941±0.002}} & \textit{\textbf{0.075±0.001}} & \textit{\textbf{0.851±0.003}} & \textit{\textbf{0.909±0.001}} & \textit{\textbf{0.129±0.002}} & \textit{\textbf{0.843±0.004}} \\
                                 &                                 & \textit{Phase}     & \textit{\textbf{0.429±0.01}}  & \textit{\textbf{0.935±0.006}} & \textit{\textbf{0.463±0.01}}  & \textit{\textbf{0.47±0.009}}  & 0.983±0.004                   & \textit{\textbf{0.501±0.009}} & \textit{\textbf{0.443±0.01}}  & 0.961±0.005                   & \textit{\textbf{0.477±0.01}}  \\
                                 & \multirow{2}{*}{Fourier-PDUNet} & \textit{Magnitude} & 0.864±0.012                   & 0.864±0.012                   & 0.79±0.008                    & 0.93±0.011                    & 0.084±0.005                   & 0.826±0.006                   & 0.905±0.002                   & 0.144±0.002                   & 0.843±0.003                   \\
                                 &                                 & \textit{Phase}     & 0.405±0.013                   & 1.095±0.003                   & 0.452±0.011                   & 0.462±0.014                   & \textit{\textbf{0.963±0.01}}  & 0.498±0.012                   & 0.401±0.012                   & \textit{\textbf{0.934±0.005}} & 0.466±0.011                   \\ \bottomrule
\end{tabular}%
}    
\end{center}
\end{table*}

\begin{table*}[htbp]
\caption{The mean temperature difference in the whole volume  between the reconstructed temperature maps obtained from the different methods and the ground truth.}\label{tab:Quantativerslt}
\begin{center}
\begin{tabular}{@{}cccc@{}}
\toprule
\textbf{Type of Undersampling} & \textbf{Undersample} & \textbf{Fourier-PDNet} & \textbf{Fourier-PDUNet} \\ \midrule
1D Varden 25\%                 & 1.221±0.035          & 0.637±0.016            & 0.641±0.018             \\
2D Varden 25\%                 & 1.296±0.031          & 0.596±0.014            & 0.657±0.019             \\
2D Varden 10\%                 & 1.299±0.032          & 0.618±0.016            & 0.643±0.022             \\ \bottomrule
\end{tabular}
\end{center}
\end{table*}

\begin{table*}[htbp]
\caption{The mean temperature difference in the tumour ROI  between the reconstructed temperature maps obtained from the different methods and the ground truth.}\label{tab:ROIQuantativerslt}
\begin{center}
\begin{tabular}{@{}cccc@{}}
\toprule
\textbf{Type of Undersampling} & \textbf{Undersample} & \textbf{Fourier-PDNet} & \textbf{Fourier-PDUNet} \\ \midrule
1D Varden 25\%                 & 0.181±0.078          & 0.052±0.025            & 0.055±0.023             \\
2D Varden 25\%                 & 0.363±0.109          & 0.028±0.005            & 0.088±0.016             \\
2D Varden 10\%                 & 0.488±0.161          & 0.063±0.009            & 0.110±0.026             \\ \bottomrule
\end{tabular}
\end{center}
\end{table*}

\end{document}